\documentclass[%
 reprint,
 amsmath,amssymb,
 aps,
]{revtex4-2}
\usepackage{soul,xcolor}
\usepackage{graphicx}
\usepackage{dcolumn}
\usepackage{bm}
\usepackage{amsmath}
\usepackage{times,mathptmx,amssymb}
\usepackage{xcolor}
\usepackage{siunitx}
\usepackage{soul}

\newcommand{\VEC}[1]{\mathbf{#1}}
\newcommand{\HAT}[1]{\hat{\mathbf{#1}}}

\begin{document}
\setstcolor{red}
\raggedbottom
\title{Tension Remodeling Controls Topological Transitions in Epithelial Tissues}

\author{Fernanda P\'erez-Verdugo}
\affiliation{Department of Physics, Carnegie Mellon University, Pittsburgh, PA 15213, USA}

\author{Shiladitya Banerjee}
\email{shiladtb@andrew.cmu.edu}
\affiliation{Department of Physics, Carnegie Mellon University, Pittsburgh, PA 15213, USA}

\begin{abstract}
\noindent Cell neighbor exchanges play a critical role in regulating tissue fluidity during epithelial morphogenesis and repair. In vivo, these neighbor exchanges are often hindered by the formation of transiently stable four-fold vertices, which can develop into complex multicellular rosettes where five or more cell junctions meet. Despite their importance, the mechanical origins of multicellular rosettes have remained elusive, and current cellular models lack the ability to explain their formation and maintenance. Here we present a dynamic vertex model of epithelial tissues with strain-dependent tension remodeling and mechanical memory dissipation. We show that an increase in cell junction tension upon contraction and reduction in tension upon extension can stabilize higher-order vertices, temporarily stalling cell rearrangements. On the other hand, inducing mechanical memory dissipation via relaxation of junction strain and stress promotes the resolution of higher-order vertices, facilitating cell neighbor exchanges. We demonstrate that by tuning the rates of tension remodeling and mechanical memory dissipation, we can control topological transitions and tissue material properties, recapitulating complex cellular topologies seen in developing organisms.
\end{abstract}
\maketitle

\section{Introduction}

Fluidization of epithelial tissues plays a vital role in coordinating large-scale structural changes in early development \cite{mongera2018fluid,saadaoui2020}, wound healing \cite{tetley2019tissue}, and collective cell migration~\cite{park2015,atia2018,barriga2018}. While multiple cell-level mechanisms contribute to tissue fluidity, including cell migration, division, and death~\cite{ranft2010}, cell neighbor exchanges are one of the most common drivers of tissue fluidity during morphogenesis~\cite{tetley2018,lemke2021dynamic}. During a neighbor exchange process occurring via a {\it T1 transition}, two cells in contact shrink their shared junction to a single point, forming a 4-fold vertex. This 4-fold vertex then extends into a new intercellular junction in a direction orthogonal to the contracting junction. While neighbor exchange processes rely on the instability of 4-fold vertices, {\it in vivo} experiments showed that 4-fold vertices can be stable for long times in developing tissues~\cite{classen2005hexagonal,blankenship2006multicellular, jessica2016local, finegan2019tricellular}. In particular, during axis elongation in \textit{Drosophila}, vertices shared by 4 or more cells (termed as {\it rosettes}), could persist for upto 15-40 min \cite{blankenship2006multicellular,finegan2019tricellular}, stalling cell neighbor exchanges. 

Experimental observations of controlled cell neighbor exchanges contrast with existing vertex models of epithelial tissues~\cite{fletcher2014,alt2017}, where stationary 4-fold vertices do not naturally arise and are energetically unstable~\cite{spencer2017vertex}.
Furthermore, experiments showed that 4-fold and higher-order vertices often restore the original cell junction, resulting in a reversible T1 process~\cite{jessica2016local,finegan2019tricellular,curran2017myosin}. By contrast, most theoretical studies treat the creation and resolution of 4-fold vertices as an instantaneous and unidirectional event triggered by junctions contracting below a length threshold \cite{comelles2021epithelial} or if neighbor exchange is energetically favorable~\cite{staddon2018,yamamoto2022non}. Others have engineered the formation of higher-order vertices by ad-hoc rules. For instance, Farhadifar et al. \cite{farhadifar2007influence,farhadifar2009dynamics} enforced the creation of 4-fold vertices by joining proximal three-fold vertices, and stalling their subsequent resolution. On the other hand, there have been recent theoretical efforts to understand the impact of non-instantaneous resolution of 4-fold vertices and probabilistic T1 events \cite{yan2019multicellular, das2021controlled, erdemci2021effect,finegan2019tricellular}. However, these studies impose the stalling of T1 events by ad-hoc rules, and they did not naturally arise from the underlying mechanics of the tissue.

To explain the physical origin of four-fold vertex stability and controlled T1 transitions in epithelial tissues, 
we extended the existing framework of vertex models~\cite{farhadifar2007influence,fletcher2014,alt2017} to incorporate dynamic tension remodeling and mechanical memory dissipation. In particular, the tension in intercellular junctions evolves in time due to changes in junctional strain above a threshold or in response to active fluctuations. We show that tension remodeling and mechanical memory dissipation leads to controlled cell neighbor exchanges, such that T1 transitions are stalled when they are not energetically favorable. By tuning the rates of tension remodeling we can control the probability of reversible and irreversible T1 transitions, as well as the timescale of stalling of 4-fold vertices. While the mechanical stability of higher-order vertices relies on the ability of cellular junctions to remodel their tension in response to strain, their resolution requires timely dissipation of mechanical memory in the system. Therefore, transient stabilization of $n$-fold vertices ($n>3)$ relies on mechanical memory dissipation, which could occur via relaxation of junctional tension, strain or noise-induced tension fluctuations. In addition to regulating tissue topology and cell morphologies, tension remodeling rates also control the emergent material properties of the tissue. In particular, we show that by tuning the rates of tension remodeling, epithelial tissues can transition between solid and fluid-like phases with tunable rates of energy dissipation. Taken together, our theory and simulations uncover the mechanical requirements for controlled T1 transitions in epithelial tissues, as well as elucidate the mechanics underlying solid-fluid transitions in epithelia.

\section{Vertex Model with tension remodeling}
\subsection{Forces and equations of motion}
To describe the dynamics of topological transitions in confluent tissues we use the framework of the vertex model \cite{nagai2001dynamic, farhadifar2007influence, staple2010mechanics, fletcher2014}, where each cell is modeled as a two-dimensional polygon, with edges representing the cell-cell junctions and the vertices representing multi-cellular junctions. The overdamped dynamics of vertices are determined by a balance of forces between friction, cell elasticity and active forces acting at intercellular junctions. Position $\VEC{r}_i$ of vertex $i$ evolves in time as
\begin{equation}
\mu \frac{{\rm d}\VEC{r}_i}{{\rm d}t}=-\frac{\partial E_{\text{el}}}{\partial \VEC{r}_i} +  \VEC{F}_i^{\text{act}}\;,
\label{eq.eq1}
\end{equation}
where $\mu$ is the vertex friction coefficient, $E_\text{el}= \sum_\alpha (K/2) \left(A_\alpha-A_{\alpha}^0\right)^2$ penalizes changes in the area $A_\alpha$ of cell $\alpha$, with respect to its target value $A^{0}_\alpha$, and $K$ is the bulk elastic modulus. Active forces arise from actomyosin contractility $\Gamma_a$ in the cell cortex, and tension $T$ at intercellular junctions, such that $\VEC{F}_i^{\text{act}}= - \sum_{\langle ij\rangle}\left[T_{ij} + \Gamma_a l_{ij}\right]\left(\partial l_{ij} /\partial \VEC{r}_i\right)$, where $T_{ij}$ is the tension on an edge connecting vertices $i$ and $j$ with length $l_{ij}$~\cite{staddon2019mechanosensitive,krajnc2021active}. Tension due to actomyosin contractility is proportional to junction length, qualitatively similar to perimeter-dependent contractility term in classical vertex models~\cite{fletcher2014}. This captures the positive feedback effect that myosin recruitment increases with increasing junction length~\cite{gustafson2022patterned}. Note that the force due to $\Gamma_a$ could also be interpreted as a conservative force arising from an energy term $\sum_{\langle ij\rangle}\Gamma_a l_{ij}^2/2$.

Several recent studies provided evidence that tension in epithelial cell junctions is not static but a dynamic quantity maintained by mechanochemical feedback processes~\cite{curran2017myosin,staddon2019mechanosensitive,cavanaugh2020rhoa,nishizawa2022,iyer2019,cavanaugh2022,khalilgharibi2019stress,clement2017}. We therefore model junctional tension as $T_{ij}(t) = \Lambda_{ij} (t)+\Delta \Lambda_{ij}(t)$, where $\Lambda_{ij}(t)$ is the deterministic part of the tension and $\Delta \Lambda_{ij}(t)$ represents stochastic fluctuations in tension. The dynamics of $\Lambda_{ij}$ is dependent on the junctional strain $\epsilon_{ij}=(l_{ij}-l_{ij}^0)/l_{ij}^0$, where $l_{ij}^0$ is the junction rest length. Tension $\Lambda_{ij}$ evolves in time as,
\begin{equation}
\frac{ {\rm d} \Lambda_{ij}}{{\rm d}t} = -\alpha(\epsilon_{ij}) (l_{ij}-l_{ij}^0) - \frac{1}{\tau_\Lambda}\left(\Lambda_{ij}-\Lambda_0\right)\;,
\label{eq.tensionrem}
\end{equation}
where the first term describes strain-dependent tension remodeling, as recently introduced by us~\cite{staddon2019mechanosensitive} and tested experimentally~\cite{cavanaugh2020rhoa,cavanaugh2020,nishizawa2022}, and the second term describes tension relaxation to a mean value $\Lambda_0$, which occurs over a longer characteristic timescale $\tau_\Lambda$. Motivated by recent experiments on single junction mechanics~\cite{cavanaugh2020rhoa,cavanaugh2020}, the rate of tension remodeling $\alpha$ (units of force per unit length per unit time) is defined as: $\alpha(\epsilon_{ij}<-\epsilon_c)=k_C$, $\alpha(\epsilon_{ij}>\epsilon_c)=k_E$, and $\alpha=0$ otherwise, where $\epsilon_c$ is a threshold strain for junction remodeling.
With positive $k_E$ and $k_C$, there is a negative feedback effect such that tension increases upon contraction at a rate $k_C$ and reduces upon stretch at a rate $k_E$, consistent with experimental observations~\cite{cavanaugh2020rhoa,khalilgharibi2019stress,nishizawa2022}. The threshold strain is motivated by optogenetics data on single junction activation that show cellular junctions only remodel their length above a threshold contraction~\cite{staddon2019mechanosensitive}. Tension remodeling above a critical strain threshold allows for irreversible junction deformation for sufficiently strong or sustained force~\cite{cavanaugh2020rhoa}. 

Additionally, cellular junctions continuously relax strain at a rate $k_L$ such that the junction rest length approaches current length as 
\begin{equation}
\frac{{\rm d}  l_{ij}^0}{{\rm d} t}= -k_L (l_{ij}^0-l_{ij})\;.
\label{eq.strainrelaxation}
\end{equation}
Strain relaxation via rest length remodeling~\cite{odell1981,munoz2013,staddon2019mechanosensitive} is a natural consequence of turnover in strained actomyosin networks~\cite{mcfadden2017}, where deformed filaments are replaced by unstrained ones.
An important consequence of strain relaxation is that memory of prior deformations are erased over a timescale $k_L^{-1}$, such that long periods of contractions can remodel junctions only upto a limit, while pulsatile contractions with periods of rest enables irreversible deformations via ratcheting~\cite{staddon2019mechanosensitive, cavanaugh2020rhoa}. 

Finally, tension fluctuations $\Delta \Lambda_{ij}$ evolve according to an Ornstein-Uhlenbeck process as~\cite{curran2017myosin,tetley2019tissue},\begin{equation}\label{eq:fluc}
\frac{{{\rm d} \Delta \Lambda_{ij}}}{{\rm d}t} = -\frac{1}{\tau}\Delta\Lambda_{ij}+\sqrt{2\sigma^2/\tau}\ \xi_{ij}(t)\;,
\end{equation}
where $\sigma$ is the fluctuation amplitude, $\xi_{ij}(t)$ is a white Gaussian noise satisfying $\langle \xi_{ij}(t) \xi_{mn}(t')\rangle = \delta(t-t')\delta_{im}\delta_{jn}$, and $\tau$ is the persistence time of tension fluctuations. Our model thus considers three principal mechanisms for erasing memory of prior mechanical state, via tension relaxation at a rate $\tau_\Lambda^{-1}$, tension fluctuations of amplitude $\sigma$, and continuous strain relaxation at a rate $k_L$. As shown later, the transient stabilization of higher-order vertices is crucially dependent on the ability of tissues to dissipate mechanical memory.

\subsection{Mechanical stability and viscoelasticity of cell junctions}
\begin{figure}[t]
 \includegraphics[width=\columnwidth]{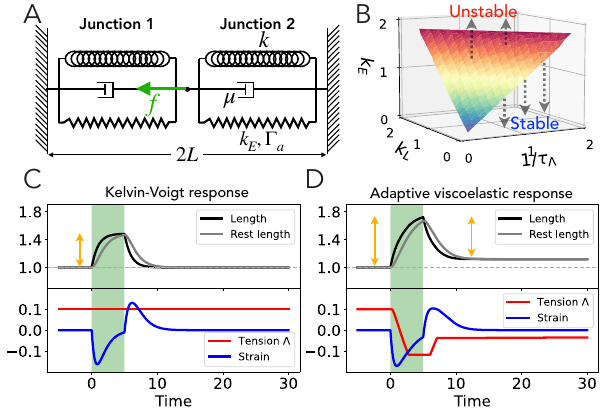} 
\caption{{\textbf{Junction stability and adaptive viscoelastic response.} (A) Schematic of a simplified two-junction model under active contraction. (B) Phase diagram in $(k_E=k_C,k_L,\tau_\Lambda^{-1})$ plane, showing the critical surface separating stable and unstable regimes of the system. Color, from blue to red, represents the value of $k_E$. (C-D) Dynamical response of junction (2) length, rest length, strain and tension $\Lambda$ for (C) $k_E=k_C=0$, and (D) $k_E=k_C=0.5$. Parameters in panels (B-C-D): $\Gamma_a=0.03$, $k_L=1$, $\tau_\Lambda^{-1}=0.03$}.}
\label{fig.Fig1}
\end{figure}

 We begin by examining the mechanical response of individual junctions to contractile forces. To do this, we simplify the model by considering a one-dimensional variant of Eqs. \eqref{eq.eq1}-\eqref{eq.strainrelaxation}, neglecting any stochastic fluctuations (as depicted in Fig.~\ref{fig.Fig1}). This simplified model focuses on a two-junction system with varying lengths, denoted as $l_1(t)$ and $l_2(t)$, respectively. Each junction unit is comprised of an elastic element with a spring constant $k$ and natural length $L$, connected in parallel to a dashpot with friction coefficient $\mu$ (Fig. \ref{fig.Fig1}A).  Additionally, an active elastic element with a rest length $l^0_{1,2}$ and contractility $\Gamma_a$ is connected in parallel to the dashpot and the spring. As previously described, the tension in the junction remodels at a rate $k_E$ under contraction and stretching, with an initial value of $\Lambda_0$, and a rest length $l^0$ that relaxes toward the current junction length with a rate $k_L$. In our analysis, we assume fixed boundary conditions, allowing only the middle vertex to move under applied forces.

To examine the mechanical stability of these junctions under applied forces, we derive a linearized system of equations for a perturbation $\delta \VEC{X} = (\delta \Gamma_1, \delta \Gamma_2, \delta l_1^0, \delta l_2^0, \delta l_1)$ around the steady-state as: $\delta \VEC{\dot X} =\mathbb{A} \delta \VEC{X}$. $\delta \VEC{X}$ follows the dynamics described in Eqs.\eqref{eq.eq1}-\eqref{eq.strainrelaxation}, with the elastic energy given by: $E_\text{el} = \frac{k}{2}(L-l_1)^2+\frac{k}{2}(L-l_2)^2$. We non-dimensionalize force scales by $kL$, length scales by $L$, and time scales by $\mu/k$, setting $k=L=\mu=1$. In this particular analysis, we assumed that $\epsilon_c=0$, such that even the slightest perturbation would induce junction tension remodeling at a rate $k_E$. We then numerically diagonalized the stability matrix $\mathbb{A}$ for different values of the rates $k_L$ (rest length relaxation), $1/\tau_\Lambda$ (tension relaxation), and $k_E$ (tension remodeling). We found that the system is stable (maximum eigenvalue of $\mathbb{A}$ $\leq0$) in the absence of tension remodeling. However, it becomes unstable at a critical value of $k_E=k_E^*$, where $k_E^*$ increases in conjunction with both $k_L$ and $1/\tau_\Lambda$ (Fig. \ref{fig.Fig1}B). Physically, this instability would manifest as junction collapse.

The one-dimensional junction model reveals adaptive viscoelastic properties that are essential for understanding tissue-level mechanical response. In addition to elasticity and dissipation through friction, cell junctions have additional sources of dissipation through tension remodeling, tension relaxation and strain relaxation. We therefore sought to analyze the viscoelastic response of individual junctions by performing a load-controlled tension test. Specifically, we applied a constant tension $f$ in the middle vertex, for a time period of $5[\mu/k]$ (Fig. \ref{fig.Fig1}C-D), and monitored the dynamics of tension and length in Junction-2, both with (Fig. \ref{fig.Fig1}C) and without tension remodeling (Fig. \ref{fig.Fig1}D). In the absence of tension remodeling ($k_E=k_C=0$), we obtain $f=(1+2\Gamma_a)\tilde{\epsilon}_2 + {\rm d}  \tilde{\epsilon}_2/{\rm d} t$ , with $\tilde{\epsilon}_2 = l_2(t)-1$. Hence, the system behaves like a Kelvin-Voigt viscoelastic solid. When $k_E>0$ the response during load is amplified, while the unloading behaviour is dependent on tension relaxation rate $\tau_\Lambda^{-1}$. For $1/\tau_\Lambda\ll \mu/k$, the relaxation during unloading is slow, leading to a steady-state with a longer equilibrium junction length $l_2>L$ (Fig.\ref{fig.Fig1}D). For $1/\tau_\Lambda< \mu/k$, we observe an undershoot in the length dynamics ($l_2(t)$) during recovery from load (figure not shown). For $1/\tau_\Lambda\gg \mu/k$, the system responds like a Kelvin-Voigt viscoelastic solid.

\subsection{Implementation of T1 transitions}

With the model mechanics defined above, we now turn to describing the dynamics governing T1 topological transitions. To simulate a T1 transition, when a junction connected by two 3-fold vertices becomes shorter than a threshold length $l_{T_1}$, one of the vertices is removed while the other is transformed into a 4-fold vertex, sustained by four shoulder junctions. During this process, each shoulder junction gains one-fourth of the deleted junction tension, and conserve it until the 4-fold vertex is resolved~\cite{sknepnek2021generating}. The latter is motivated by experimental observations of Myosin-II accumulation around junctions proximal to 4-fold vertices \cite{curran2017myosin}. We then create a new junction of length $l_{\text{birth}}=1.5l_{T_1}$, tension $\Lambda_\text{birth}\sim \Lambda_0+\Gamma_a l_{\text{birth}}$, and attempt to resolve the 4-fold vertex in two different directions -- one along the original contracting junction (resulting in reversible T1) and the other approximately orthogonal to it (leading to neighbor exchange), as experimentally observed \cite{classen2005hexagonal,curran2017myosin}. To decide the final resolution configuration we follow an approach previously introduced in Ref.~\cite{spencer2017vertex}. If the force between the vertices of the newly created junction is attractive, then the 4-fold is considered stable and the T1 transition is stalled. Otherwise, the 4-fold vertex resolves along the direction with the largest separation force~\cite{spencer2017vertex, curran2017myosin, duclut2022active}, resulting in reversible or irreversible T1 transitions (see Supplemental Material for details). As discussed later, the specific choice of T1 transition parameters, such as $\Lambda_\text{birth}$ and $l_\text{birth}$, as well as the choice of tension resetting rule does not influence our main conclusions on the formation and stability of 4-fold vertices.\\

\noindent {\it Higher-order vertex formation and resolution.--}
The rules for 4-fold vertex formation, as described above, can also be applied to the merging of a stable $n$-fold and a 3-fold vertex, allowing the possibility of a $(n+1)$-fold vertex in the tissue. During the creation of such a $(n+1)$-fold vertex, each shoulder junction gain $1/(n+1)$-th of the tension of the deleted junction, as in the $n=3$ case previously described. However, a more general idea of the resolution directions is needed. Here we propose that a $(n+1)$-fold vertex can be resolved into a $n$-fold and a 3-fold vertex in $(n+1)$ posible directions given by $\left(\VEC R_c^{\alpha} - \VEC r_{n+1}\right)/\mid \VEC R_c^{\alpha} - \VEC r_{n+1}\mid$, where $\VEC R_c^{\alpha}$ is the center of one ($\alpha$) of the $(n+1)$ cells surrounding the $(n+1)$-fold vertex with position $\VEC r_{n+1}$. See Supplemental Material for further details on $n$-fold vertices, Fig.~S2, and Movie 4 for a simulated tissue in which 3-fold, 4-fold and 5-fold vertices are allowed. In the rest of this manuscript, we allow for only 3-fold and 4-fold vertices.

\begin{figure}[t]
 \includegraphics[width=\columnwidth]{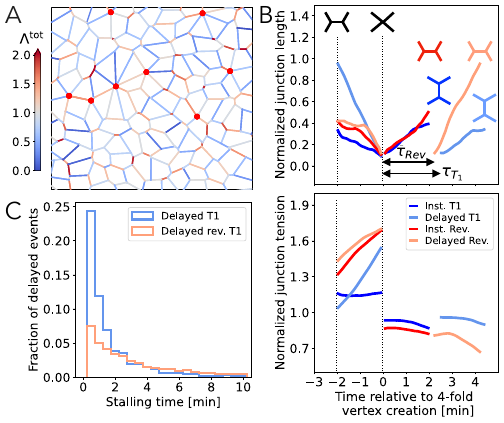} 
\caption{{\bf Delayed T1 transitions and intercalation dynamics in remodeling tissues}. (A) Representative section of a simulated epithelial tissue with tension remodeling ($k_C/k_L=0.2$, $k_E/k_L=0.17$) at $t\sim\SI{6}{\hour}$. Red circles represent 4-fold vertices, and colored cell edges represent the total tension, in units of $\Lambda_0$. (B) Normalized junction length (top) and tension (bottom), as a function of time relative to the 4-fold vertex creation, for instantaneous and delayed T1 events. (C) Histogram of the stalling time for delayed irreversible T1 (blue) and delayed reversible T1 (red) events. }
\label{fig.Fig2}
\end{figure}

\section{Results}
\subsection{Tension remodeling controls T1 transitions} 
To characterize the role of tension remodeling on T1 transitions, we first simulated a disordered tissue comprising $\sim$500 cells in a box with periodic boundary conditions. In simulations, we non-dimensionalized force scales by $K(A_\alpha^0)^{3/2}$, length scales by $\sqrt{A_\alpha^0}$, time scales by $\mu/KA_\alpha^0$, setting $K=1$, $\langle A_\alpha^0\rangle=1$, and $\mu=0.2$ ($\sim\SI{28}{\second}$), where $\langle ..\rangle$ represents population average. 
The initial state of the simulations is characterized by having zero initial junction strain ($l_{ij}=l_{ij}^0$), $\langle l_{ij}^0 \rangle \sim 0.62$ and $\langle \Lambda_{ij}\rangle=\Lambda_0=0.1$, which is also the mean value for the tension of a newly created junction (see Supplemental material, Fig.~S1). We let the tissue evolve from an energy relaxed state with chosen values of active contractility $\Gamma_a$, active fluctuations of amplitude $\sigma$, threshold strain $\epsilon_c=0.1$, strain relaxation rate $k_L$, with different values tension remodeling rates $k_E$ and $k_C$. A representative tissue snapshot is shown in Fig.~\ref{fig.Fig2}A, for a particular simulation using $k_C/k_L=0.2$ and $k_E/k_L=0.17$, which displays multiple (transiently stable) 4-fold vertices (red circles) representing stalled T1 transitions (Movie 1). 

Four different types of dynamics are observed during T1 processes (Fig.~\ref{fig.Fig2}B, Movie 1): instantaneous irreversible T1, delayed irreversible T1 with a stalled 4-fold vertex, instantaneous reversible, and delayed reversible T1 events. In all these cases, tension increases during contraction prior to 4-fold vertex formation, as a consequence of tension remodeling. Subsequently, tension decreases via remodeling after the 4-fold vertex resolves into an extending junction (Fig.~\ref{fig.Fig2}B). 
Tension remodeling can decrease the local tensions in stretched shoulder junctions, promoting 4-fold stabilization. Specifically, a stalled 4-fold vertex arises when $f_\text{4fold}=\left(\VEC f_i - \VEC f_j\right)\cdot \HAT r_{ij}< 2\Lambda_{\text{birth}}$, where $\HAT r_{ij}=\left(\VEC r_i-\VEC r_j\right)/\mid \VEC r_i-\VEC r_j\mid$, and $\VEC f_i$ and $\VEC f_j$ are the forces acting on the two tricelullar vertices, $i$ and $j$, created in the attempt of 4-fold vertex resolution. These forces arise from tensions in the shoulder junctions as well as pressures in the neighboring cells resisting changes in cell area.
When the local tension increases due to strain-driven remodeling or contractility, vertex stability is lost, resulting in delayed T1 or a delayed reversible event (Fig.~\ref{fig.Fig2}B). Fig.~\ref{fig.Fig2}C shows the distribution of stalling times, for both reversible and irreversible T1 events, suggesting that some 4-fold vertices can be resolved near-instantaneously, while others can remain stalled for longer periods. Without tension remodeling ($k_E=k_C=0$), we recover the standard vertex model where T1 transitions occur instantaneously and 4-fold vertices are unstable (Movie 2). For negative values of the tension remodeling rates, we obtain a model of positive feedback between tension and strain, where T1 transitions are observed to occur instantaneously (see Supplemental Material, Fig.~S3).

\begin{figure}[t]
 \includegraphics[width=1\linewidth]{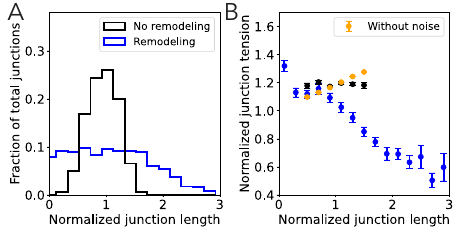} 
\caption{{\bf Tension remodeling promotes asymmetric length distribution}. (A) Histogram of junction length at steady-state, for a tissue with remodeling (blue, $k_C/k_L=0.2$ and $k_E/k_L=0.17$) and without remodeling (black, $k_C=k_E=0$). (B) Correlation between total junction tension (normalized) and junction length (normalized) in tissues with (blue) and without (black) junction remodeling. The orange data points show the positive correlation between deterministic tension (i.e., junction tension without the fluctuating part) and junction length, in the absence of junctional tension remodeling. Error bars represent $\pm$1 standard error of mean.}
\label{fig.Fig3}
\end{figure}

Dynamics of the model tissue with tension remodeling, as characterized in Fig.~\ref{fig.Fig2} (Movie 1), settle into a fluctuating steady-state with an asymmetric distribution of junction length (Fig.~\ref{fig.Fig3}A), as observed in mature {\it Drosophila} epithelium~\cite{curran2017myosin}. 
Furthermore, a negative correlation is observed between junction length and tension in the fluctuating steady-state (Fig.~\ref{fig.Fig3}B), analogous to the negative correlation between junction length and myosin intensity seen experimentally~\cite{bardet2013pten, curran2017myosin}. By contrast, without tension remodeling ($k_E=k_C=0$), junction length distribution is symmetric (Fig.~\ref{fig.Fig3}A). In this case, junction length is positively correlated with the deterministic part of the tension ($\Lambda_0 + \Gamma_a l_{ij}$) (orange dots in Fig.~\ref{fig.Fig3}B). However, this positive correlation is lost when we consider the total junction tension, including the fluctuating part, since the amplitude of the tension fluctuations ($\sigma$) is comparable to $\langle \Gamma_a l_{ij}\rangle$ (black dots in Fig.~\ref{fig.Fig3}B).

\subsection{Stability of 4-fold vertices }  
During a T1 transition, 4-fold vertex can be transiently stable if the tension in the extending shoulder junctions are small compared to the tension in the newly created junction. This can be achieved via tension remodeling in the extending shoulder junctions, controlled by the rate $k_E$.
To understand how $k_E$ affects vertex stability, we studied an effective mean-field model consisting of symmetric cell junctions embedded in an effective elastic medium (see Supplemental Material, Fig.~S4). We activated contraction in chosen junctions by increasing $\Gamma_a$. During this process, the contracting (extending) junctions increase (decrease) their tension at a rate $k_C$ ($k_E$). We found that if $\beta k_E>k_C$, the global tissue tension decreases, promoting mechanical stability of the 4-fold vertex, where $\beta$ is the ratio of the total length gained by the extending junctions to the total length lost by the contracting junctions.
\begin{figure}[t]
 \includegraphics[width=\columnwidth]{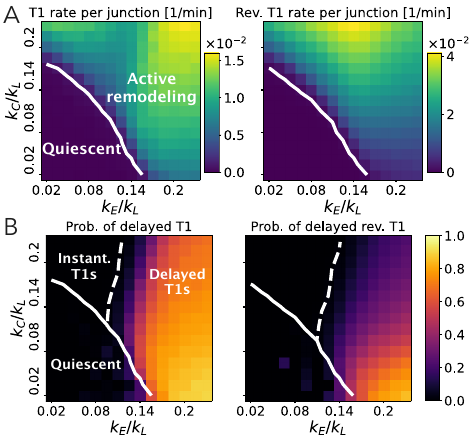} 
\caption{{\bf Junction tension remodeling regulates T1 transition rates}. (A) Rates of T1 (left) and reversible (right) transitions for different values of $k_E/k_L$ and $k_C/k_L$. Solid lines represent $10^{-3}$ T1 events per junction per minute. (B) Probability of delayed irreversible T1 transitions (left) and reversible T1 events (right), for different values of $k_E/k_L$ and $k_C/k_L$. Dashed line represents $1\%$ probability.}
\label{fig.Fig4}
\end{figure}

To further investigate the role of tension remodeling on T1 transitions, we performed numerical simulations using different values of the tension remodeling rates, $k_C/k_L$ and $k_E/k_L$ in the range [0.02,0.23] (Fig.~\ref{fig.Fig3}). From fits to experimental data on single junction deformations, it was determined that $k_C/k_L\approx 0.14$~\cite{staddon2019mechanosensitive,cavanaugh2020rhoa} and $k_E/k_L\approx 0.12$~\cite{nishizawa2022}.
We found that the tension remodeling rates controlled the rate of T1 transitions as well as the probability of delayed T1 transitions. For very small values of $k_E$ and $k_C$, the tissue is in a quiescent state, where T1 events are scarce ($<10^{-4}$ per junction per minute) and occur instantaneously (Fig.~\ref{fig.Fig3}A). Due to the lack of appreciable tension remodeling in the quiescent state, tensions in the shoulders of an intercalating junction $\sim \Lambda_0$, with negligible resistive pressure in the surrounding cells since $A_\alpha\sim A_\alpha^0$. 
As a result, $f_\text{4fold}$ remains larger than $2\Lambda_\text{birth}$, making the 4-fold vertex unstable.
For larger values of $k_C$ and $k_E$, irreversible and reversible T1 events are mainly driven by tension remodeling, inducing wider pressure distributions (Fig.~S5), and higher rates of T1 events (Fig.~\ref{fig.Fig4}A). In this parameter regime, $f_\text{4fold}$ depends on both tensions and pressures in the surrounding cells. Since tension remodeling dynamics is fast compared to pressure relaxation, pressure-like forces make $f_\text{4fold}$ larger in the original direction of contraction, turning the reversible T1 events more probable. In the presence of tension remodeling, T1 events either occur instantaneously or are delayed, with probabilities given in Fig.~\ref{fig.Fig4}B. The probabilities of delayed events, and hence the presence of stalled 4-fold vertices, depend strongly on $k_E$, as predicted analytically. 4-fold stability increases for large $k_E$ and small $k_C$, reaching stalling times of $5$ min on average (Fig.~S6), consistent with experimental data~\cite{finegan2019tricellular}. 
\begin{figure}[t]
 \includegraphics[width=\columnwidth]{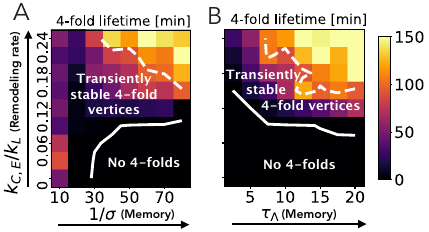} 
\caption{{\bf Mechanical memory dissipation promotes T1 transitions}. Maximum lifetime of 4-fold vertices, as a function of tension remodeling rate $k_C/k_L$ (with $k_C=k_E$) and the regulators of mechanical memory: (A) inverse of noise amplitude $1/\sigma$ (with $\tau_\Lambda=10$), and (B) tension relaxation timescale $\tau_\Lambda$ (with $1/\sigma=50$). The dashed contour represents $\SI{100}{\min}$ lifetime. Below the solid line, there are less than $10^{-3}$ 4-fold vertices resolved per junction per minute. }
\label{fig.Fig5}
\end{figure}

\begin{figure*}[t]
 \includegraphics[width=\linewidth]{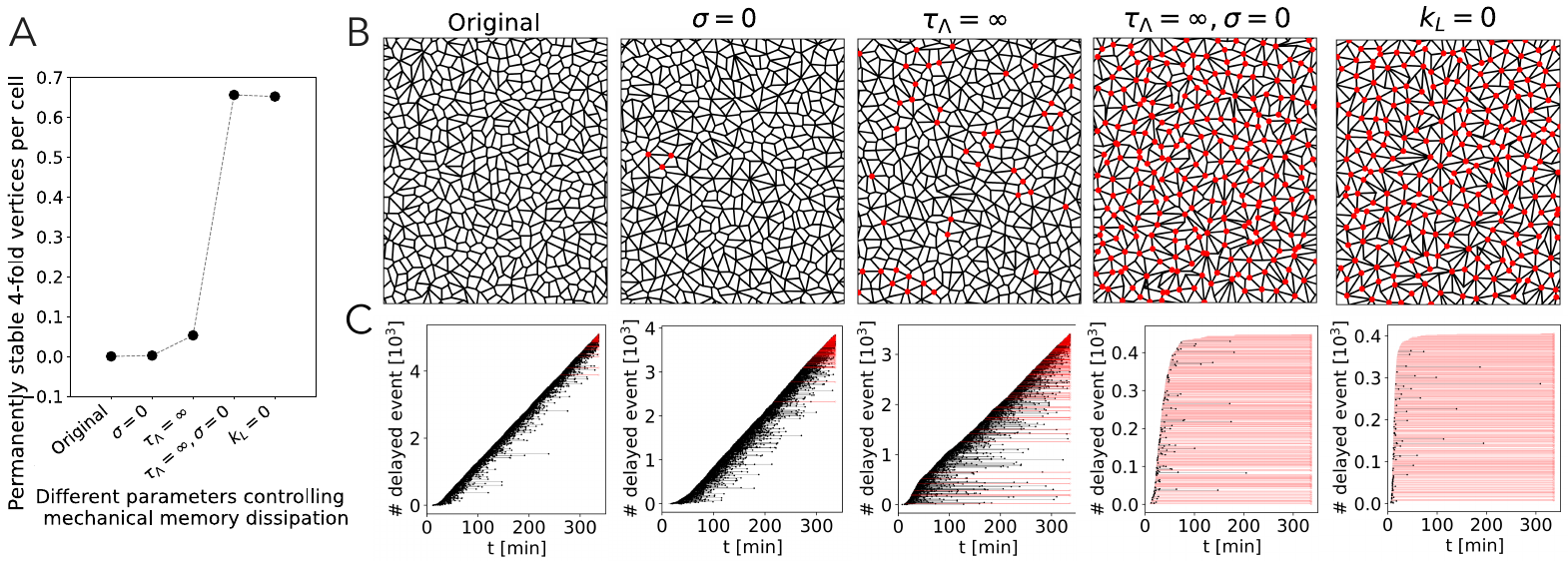} 
\caption{\textbf{ Transient stability of 4-fold vertices relies on mechanical memory dissipation}. (A) Permanently stable 4-fold vertices per cell, for an active tissue with $k_C=0.1,k_E=0.2$, and for different parameters controlling mechanical memory dissipation, in the following order (left to right): i) Original simulation with all modes of memory dissipation, with finite $\sigma$, $k_L\neq0$, and $\tau_\Lambda^{-1}\neq0$; ii) No tension fluctuations, with $\sigma=0$, $k_L\neq0$, and $\tau_\Lambda^{-1}\neq0$; iii) No tension relaxation, with finite $\sigma$, $k_L\neq0$, and $\tau_\Lambda^{-1}=0$; iv) No tension fluctuations and tension relaxation, with $\sigma=0$, $k_L\neq0$, and $\tau_\Lambda^{-1}=0$; v) No strain relaxation, with finite $\sigma$, $k_L=0$, and $\tau_\Lambda^{-1}\neq0$. (B) Tissue configurations showing the steady-state morphology (at $\sim \SI{350}{\min}$), where red solid circles represent 4-fold vertices that have been stable for more than $\SI{100}{\min}$ by the end of each simulation. Simulations with strain relaxation consider $k_L=1$. (C) Number of delayed T1 events vs time, corresponding to the simulated tissues shown in panel (B). Each horizontal line represents the creation of a 4-fold vertex. Black lines represent 4-fold vertices that are resolved through the simulation (at the time highlighted by a black dot), over a timescale longer than 6 seconds. Red lines that finish in an empty edge-colored red circle, represent 4-fold vertices that are not resolved during the simulation.}
\label{fig.Fig6}
\end{figure*}

It is important to note that the phase diagrams for T1 rate per junction, probability of delayed T1 transitions (Fig.~\ref{fig.Fig4}), as well as the probability distributions of T1 stalling times remain qualitatively the same for different choices of strain threshold parameter $\epsilon_c$, length of newly created junctions $l_{\text{birth}}$, and tension resetting rules upon T1 transitions (See supplemental material, Figs.~S7-S12).

\subsection{Mechanical memory dissipation promotes T1 transitions} 
Our study so far demonstrates that tension remodeling rates control the probability of T1 events as well as the stability of 4-fold vertices. Since higher-order vertices appear transiently in living tissues~\cite{blankenship2006multicellular, jessica2016local, finegan2019tricellular,curran2017myosin}, we wondered what tissue properties would regulate the lifetime of 4-fold vertices. To that end, we found that T1 stalling time increases with both the inverse of the noise magnitude $1/\sigma$, and the timescale of stress relaxation $\tau_\Lambda$ (Figs.~\ref{fig.Fig5}, S11, S12). 
On the contrary, the rate of T1 events increases with $\sigma$~\cite{curran2017myosin,tetley2019tissue,kim2021}, and decreases with $\tau_\Lambda$ (Figs.~S13-S14). For low or no noise ($\sigma=0$), tension fluctuations are diminished with 4-fold vertices being present for more than $\SI{100} {\ \minute}$ (see Figs.~\ref{fig.Fig5}A, \ref{fig.Fig6}C and Movie 3). Interestingly, for very high $\sigma$ (limit of no mechanical memory), tissues can reach certain geometrical configurations that allow the existence of stalled 4-fold vertices even in the absence of tension remodeling (Figs.~\ref{fig.Fig5}A). However, in such instances the stalling time cannot be dynamically controlled. For intermediate levels of noise, active tension remodeling induces 4-fold vertex formation with controllable lifetime. 
Additionally, without tension relaxation ($\tau_\Lambda=\infty$) or strain relaxation ($k_L=0$), tissues develop permanently stable 4-fold vertices (Fig.~\ref{fig.Fig6}). Particularly, for the cases $k_L=0$ and $\lbrace\tau_\Lambda=\infty, \sigma=0\rbrace$, the system quickly gets stuck in geometrical configurations with high density of stable 4-fold vertices. Our model also leads to mechanical memory dissipation via tension resetting during a 4-fold vertex resolution into 3-fold vertices. We find that a persistent-tension rule during a T1 transition also leads to permanently stable 4-fold vertices (see Fig.~S15), as seen experimentally in \textit{Drosophila} (pupal wing) lacking the tumor suppressor PTEN \cite{bardet2013pten}.

Thus, transiently stable 4-fold vertices, defined by having stalling times smaller than $\SI{100}{\ \minute}$,  require two fundamental ingredients (Fig.~\ref{fig.Fig5}): (1) a negative feedback between junctional tension and strain (tension decreases with increasing strain), and (2) mechanical memory dissipation via strain relaxation ($k_L\neq 0$), tension relaxation (finite $\tau_\Lambda$) and noise-induced fluctuations (intermediate $1/\sigma$). While there are other recent models with tension-strain feedback~\cite{krajnc2021active,sknepnek2021generating,noll2017active,gustafson2022}, those do not concurrently satisfy the above two specific criteria for tension remodeling and mechanical memory dissipation, and therefore cannot capture transiently stable 4-fold vertices (see Discussion).

\begin{figure}[htp]
 \includegraphics[width=1\linewidth]{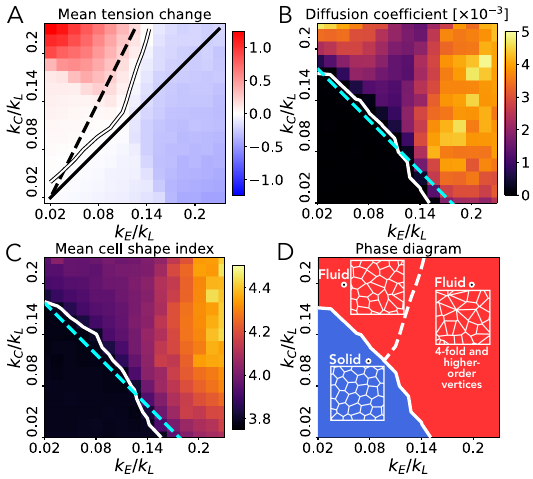} 
\caption{{\bf Emergent tissue mechanics from cell junction remodeling.} (A) Colormap of mean tension change (in units of $\Lambda_0$) at steady-state of the remodeled tissue, as functions of $k_E/k_L$ and $k_C/k_L$. Black (white) curves correspond to numerical results obtained with the effective medium  models (simulations), representing no change in tension. Dashed-black: $2k_E=k_C$ and Solid-black: $k_E = k_C$. (B) Diffusion coefficient, $D$, as functions of $k_E/k_L$ and $k_C/k_L$. The white curve represents $D=10^{-4}$ $\left[\langle A_0^\alpha \rangle ^2/ \SI{}{\minute}\right]$. (C) Mean cell shape index, $q$, as functions of the tension remodeling rates. The white curve represents $q=3.81$. Dashed cyan lines in panels B and C represent prediction of the mean-fied model, with fitted effective medium stiffness $k=0.09$. (D) Phase diagram showing transitions between solid (blue) and fluid (red) states of the tissue, with solid white curve representing the phase boundary. Below the dashed white curve in the fluid phase, stable 4-fold and higher-order vertices are prevalent.}
\label{fig.Fig7}
\end{figure}
\subsection{Tension remodeling rates control tissue material properties}
Tension remodeling rates not only control the kinetics of T1 transitions and tissue topology, but also regulates tissue material properties. To characterize mechanical properties at the tissue-level we first examined the effects of junction remodeling on average tissue tension, since low tension is associated with fluid-like tissues whereas high tension promotes solidity~\cite{bi2015,tetley2019tissue}. 
To this end, we computed the mean change in tissue tension from an initial steady-state, as functions of the tension remodeling rates, $k_E$ and $k_C$ (Fig~\ref{fig.Fig7}A). Here, the initial state is chosen as the steady-state state of the tissue with a constant mean junctional tension, $\langle \Lambda\rangle=\Lambda_0$, and zero junctional strain.
Therefore, any changes in mean tissue tension would reflect the effects of junction tension remodeling, resulting in junction length variations.

We found that the mean tension change is negative in the parameter space with non-zero probabilities of delayed T1s (Fig~\ref{fig.Fig4}B), suggesting a loss of tissue rigidity. 
The white solid line in Fig.~\ref{fig.Fig7}A represents the phase boundary obtained from simulations, where the mean tissue tension does not change. For small ($k_E,k_C$), the phase boundary follows the line $k_E=k_C$ (solid black), as predicted by the mean-field model in a system conserving the total junction length (see Supplemental Material). For large ($k_E,k_C$), the phase boundary increases in slope, as predicted in a system increasing its total junction length. In a two-dimensional mean-field model, consisting of five symmetric cell junctions subject to an increase in total junction length (see Supplemental Material), the predicted phase boundary is $2 k_E=k_C$ (dashed black line). To directly test the role of tension remodeling on tissue mechanical properties, we performed finite shear simulations (Fig.~S16). These simulations revealed that tissues with high tension remodeling, exhibiting transiently stable 4-fold vertices and a negative mean tension change, are associated with an enhanced rate of energy and stress release.

\begin{figure*}[t]
 \includegraphics[width=\linewidth]{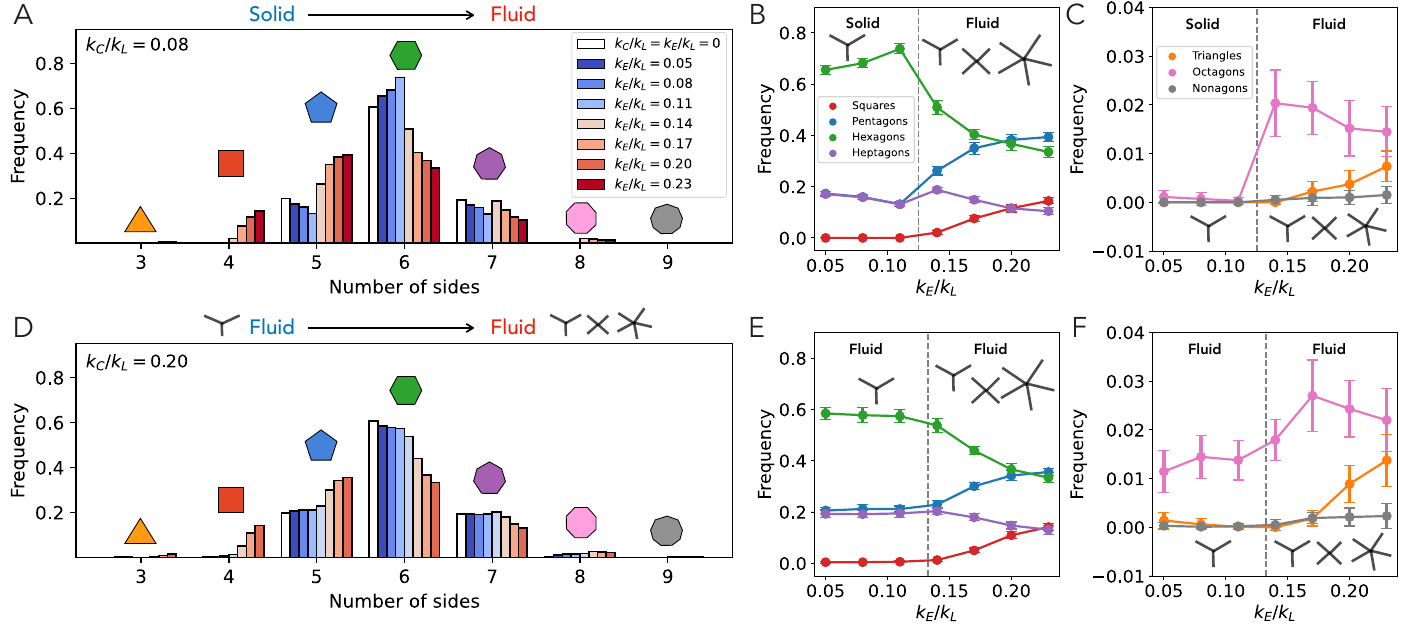} 
\caption{{\bf Tension remodeling influences the geometry of cell packing}. Mean frequency of polygon sidedness in tissues with (A-B-C) $k_C/k_L=0.08$, and (D-E-F) $k_C/k_L=0.20$, as a function of $k_E/k_L$, considering 20 random initial seeds for each simulation. White bars in panels A and D show the distribution of polygon sidedness in tissues without tension remodeling ($k_E=k_C=0$). Error bars in B-C-E-F represent $\pm$1 standard deviation.}
\label{fig.Fig8}
\end{figure*}
From the measurements of changes in tissue tension, we found that tissues with smaller values of $k_E$ and $k_C$ maintained a constant mean tension (Fig.~\ref{fig.Fig7}A), with very low rates of neighbor exchange (Fig.~\ref{fig.Fig4}A), characteristic of an arrested state. 
To quantify cell movement, we measured the mean-squared displacements of the cell centers (Supplemental Material) to compute the average diffusivity ($D$) of cells (Fig.~\ref{fig.Fig7}B). We found that cells with smaller values of ($k_E,k_C$) do not diffuse significantly ($D<10^{-4}$), representing solid-like tissues with mostly hexagonal cell shapes. Diffusivity increases with $k_E$ and $k_C$, such that the tissue is liquid-like when $(k_E+k_C)/k_L$ is larger than a critical value (Fig.~\ref{fig.Fig7}B). Interestingly, tissues possessing higher-order vertices are fluid-like with a high cell diffusivity (Fig.~\ref{fig.Fig7}B).

In vertex models describing isotropic tissues, fluidity is related to the observed cell shape index $q$~\cite{bi2015,park2015,damavandi2022universal}, defined as the mean ratio between the perimeter and the square root of the cell area. A fluid-solid phase transition occurs at $q=3.81$, such that the tissue is solid-like for $q<3.81$. The rigidity transition is related to the mechanical stability of junctions, which occurs in the mean-field theory when $(k_C+k_E)/2k_L$ is smaller than the effective medium stiffness $k$ (dashed line in Figs.~\ref{fig.Fig7}B-C, see Supplemental Material).
From our simulations, we obtained an excellent agreement between the contours $D=10^{-4}$ and $q=3.81$ (white curves in Figs.~\ref{fig.Fig7}B,C).
Our theory thus relates the fluidity of confluent tissues and their emergent topology to the rates of tension remodeling, $k_C$ and $k_E$ (Fig.~\ref{fig.Fig7}). In particular, we find that 4-fold and higher-order vertices can become stable in fluid tissues if $\beta k_E>k_C$ ($\beta>1$), such that asymmetric tension remodeling reduces mean tissue tension. 

\subsection{Tension remodeling regulates the geometry of cell packing}
In addition to controlling the frequency and timescale of T1 transitions, and tissue material properties, tension remodeling also influences the geometry of cell packing in epithelia. To characterize the cell packing geometries, we measured the fraction of cells in different polygon classes, characterized by their number of sides. Fig.~\ref{fig.Fig8} shows the distribution of number of polygon sides for different values of $(k_C/k_L,k_E/k_L)$, alongwith the polygon sidedness for tissues with no tension remodeling (white bars). 
 In tissues without tension remodeling, only pentagons, hexagons and heptagons are observed in the ground-state of the vertex model.
 We found that increasing $k_E$ in a solid tissue increases the number of hexagonal cell shapes while decreasing the relative numbers of pentagons and heptagons. The behavior is similar to what was obtained in Ref.~\cite{curran2017myosin}, when increasing the mean line tension in the cell edges. On the other hand, when increasing $k_E$ in a fluid tissue, the number of hexagons and heptagons decrease, while pentagons increase in number. In fluid tissues, triangles, squares, octagonal, and nonagonal cell shapes are also observed as $k_E$ is increased. 
 
Experimental data demonstrate the presence of diverse polygonal cell shapes, ranging from triangles to nonagons, in various tissues such as the larval wing disc of Drosophila~\cite{farhadifar2007influence,gibson2006emergence}, tail epidermis of Xenopus~\cite{gibson2006emergence}, and in the epidermis of Hydra~\cite{gibson2006emergence}. However, the origin of these irregular cell packing geometries, whether arising from cell divisions, anisotropic forces within the tissue, active tension remodeling, or a combination thereof, remains inconclusive. Here we establish that tension remodeling alone is sufficient to induce irregularities in cell packing. Future investigations that integrate cell divisions with tension remodeling and anisotropy will provide quantitative insights into the relative contributions of each of these factors in governing cell packing geometry in disordered epithelia.

\section{Discussion}
One common assumption in existing cell-based models of epithelial tissues is that epithelia resemble foam-like networks, consisting of bicellular junctions that connect tricellular vertices~\cite{farhadifar2007influence,fletcher2014,alt2017}. More complex structures, such as rosettes, where four or more junctions meet, are widely observed in vivo~\cite{blankenship2006multicellular, jessica2016local, finegan2019tricellular,curran2017myosin} but are not stable structures in existing vertex models. In this manuscript, we provide the first theoretical model for the spontaneous emergence of stable higher-order vertices and elucidate the underlying physical principles that regulate their assembly and lifetime. In particular, we identify two general physical principles that are necessary and sufficient for the formation and transient stability of higher-order vertices: (1) strain-dependent tension remodeling (specifically a negative feedback between tension and strain) and (2) mechanical memory dissipation. First, we show that the ability of cellular junctions to actively decrease tension under extension and increase tension under contraction promotes the formation of 4-fold vertices that are precursors to T1 transitions. The model for strain-dependent tension remodeling is derived from recent studies on single-junction mechanics~\cite{staddon2019mechanosensitive, cavanaugh2020rhoa}. Secondly, the relaxation of mechanical strain, tension, noise induced fluctuations enable the dissipation of mechanical memory over time, which is necessary for the timely resolution of four-fold vertices. Resolution of four-fold vertices can occur instantaneously or non-instantaneously, resulting in reversible or irreversible T1 transitions. Particularly, T1 resolution or stalling time increases with both $1/\sigma$ and $\tau_\Lambda$. For very small values of tension remodeling rates, we find that the classical vertex model results are recovered, where 4-fold vertices are unstable and resolved through T1 events. 

Our modified version of the vertex model treats each junction as an independent entity, in contrast to the classical vertex model with a perimeter-dependent contractility term. This distinction is particularly significant because the cell perimeter-dependent model introduces non-local tensions in newly formed junctions following a T1 transition, which depends on the shapes of the two adjacent cells. While there is no experimental evidence supporting this non-local tension term, its inclusion in the model can result in stabilization of four-fold vertices under specific initial conditions of cell shapes, assuming they follow similar force-dependent rules for the formation and resolution of four-fold vertices, as in our model. However, the existence of these four-fold vertices is reliant on the choice of initial geometrical conditions.

In recent years, many studies have delved into the profound impact of mechanical feedback on epithelial tissue dynamics~\cite{noll2017active,krajnc2021active,gustafson2022patterned,sknepnek2021generating}. These include studies that considered a positive feedback between tension and strain~\cite{banerjee2015,noll2017active,gustafson2022patterned,sknepnek2021generating,hino2020} or a negative feedback between tension and strain akin to our model~\cite{krajnc2021active,cavanaugh2020,nishizawa2022,staddon2019mechanosensitive}. 
However, none of these models fulfill the combined requirements of negative feedback-based tension remodeling and memory dissipation, thus falling short in capturing the transient stability of higher-order vertices within epithelial tissues. In the active tension network model introduced by Noll et al. \cite{noll2017active} and later developed in Gustafson et al. \cite{gustafson2022patterned}, tissue dynamics is governed by tension remodeling without elastic restoring elastic forces. Opposite to our work, the authors considered that tension increased in elongated junctions, and decreased in contracted junctions, which was necessary to ensure mechanical stability in the absence of elastic restoring forces. Under such rules, 4-fold vertices cannot be stabilized as the net pulling force in the extending shoulder junctions would always exceed the force between two proximal tricellular vertices. On the other hand, Krajnc et al. \cite{krajnc2021active} considered similar physical rules for active tension remodeling as in Staddon et al.~\cite{staddon2019mechanosensitive}, but did not include mechanisms for strain relaxation that we found to be necessary to avoid permanently stable 4-fold vertices (see Fig.~\ref{fig.Fig6}). A recent study by Sknepnek et al. \cite{sknepnek2021generating} implemented specific rules for mechanosensitive myosin dynamics that regulated cell junction tension. In contrast to our model, Sknepnek et al. found that the tension in the shoulder junctions of a contracting junction increases, which promote the instability of the 4-fold vertices as argued by us. Second, Sknepnek et al. did not consider relaxation and remodeling of passive tensions in the vertex model. In the absence of total tension relaxation, permanently stable 4-fold vertices would arise (see Fig.~\ref{fig.Fig6}).

We show that by tuning the values of the tension remodeling rates $k_C$ and $k_E$, the tissue can be driven through two distinct phase transitions. First, the model predicts a rigidity crossover if $(k_C + k_E)/k_L$ is smaller than a critical value, which corresponds to a mean observed cell shape index of 3.81. Below the rigidity threshold, T1 events are scarce and tissues are highly ordered, with a high fraction of  hexagonal cells, followed by smaller fractions of pentagons and heptagons. Increasing the tension remodeling rates above the rigidity threshold leads to an increase in T1 events, with a wider distribution of polygon sides (from triangles to nonagons) and an asymmetric distribution of junction length, as seen in experimental data~\cite{curran2017myosin}. Our model thus belongs to a broader class of vertex models with universal rigidity features as suggested in Ref.~\cite{damavandi2022universal}. 
Second, our model predicts a transition in tissue topology from unstable to stable 4-fold vertices in the fluid phase. In this phase, mechanical stability is reached with a lower value of mean tension, and there is asymmetry in the rates of tension remodeling in response to junction contraction and extension. Stable 4-fold vertices can also lead to the formation of even higher order vertices, as shown in Movie 4 where we allowed up to 5-fold vertices.
It is important to recall that our model assumes an isotropic tissue. In anisotropic tissues cell shape may not be a direct proxy for fluidity \cite{wang2020anisotropy}. This could explain the presence of stable 4-fold vertices in tissues with isotropic cell shapes (low shape index) as the \textit{Drosophila} pupal wing \cite{classen2005hexagonal}.

Previous studies have enforced the creation of 4-fold and higher-order vertices in canonical vertex models~\cite{yan2019multicellular} and imposed ad-hoc rules for the stalling of T1 events~\cite{das2021controlled, erdemci2021effect,finegan2019tricellular}. For instance, Finegan et al.~\cite{finegan2019tricellular} implemented probabilities for successful T1 resolution and imposed the no-resolution of rosettes, while Das et al.~\cite{das2021controlled} and Erdemeci et al.~\cite{erdemci2021effect} introduced clocks for T1 transitions. In our model, the stability of higher-order vertices is naturally linked to the mechanical state of the tissue, in particular they arise in low tension tissues. Interestingly, an increase in the mean tension in this model
does not imply a more solid-like tissue. Instead, high-tension systems are obtained in tissues with high rates of instantaneous T1 events (Supplemental Material), inducing cellular motion through these topological rearrangements and increasing diffusion, making the tissue more fluid-like. It has been previously reported that the presence of higher-order vertices leads to rigidification of tissues \cite{yan2019multicellular}. While we did not directly evaluate shear modulus of the tissue, we found that the presence of stalled 4-fold vertices reduces the rate of cell neighbor exchanges (Fig.~S17). In addition, the mechanical stability of 4-fold vertices demands a liquid-like tissue with low overall tension. This implies there are many T1 events occurring in the presence of stable 4-fold vertices, as observed during {\it Drosophila} axis elongation \cite{finegan2019tricellular}.

\acknowledgments
{\it Acknowledgements} -- We thank Michael Staddon for useful discussions and acknowledge support from the National Institutes of Health grant NIH R35-GM143042.


\begin{thebibliography}{1}

\bibitem{perez2020vertex}
F.~P{\'e}rez-Verdugo, J.-F. Joanny, and R.~Soto, ``Vertex model instabilities
  for tissues subject to cellular activity or applied stresses,'' {\em Physical
  Review E}, vol.~102, no.~5, p.~052604, 2020.

\bibitem{bi2015}
D.~Bi, J.~Lopez, J.~M. Schwarz, and M.~L. Manning, ``A density-independent
  rigidity transition in biological tissues,'' {\em Nature Physics}, vol.~11,
  no.~12, pp.~1074--1079, 2015.

\bibitem{farhadifar2007influence}
R.~Farhadifar, J.-C. R{\"o}per, B.~Aigouy, S.~Eaton, and F.~J{\"u}licher, ``The
  influence of cell mechanics, cell-cell interactions, and proliferation on
  epithelial packing,'' {\em Current Biology}, vol.~17, no.~24, pp.~2095--2104,
  2007.

\bibitem{staddon2019mechanosensitive}
M.~F. Staddon, K.~E. Cavanaugh, E.~M. Munro, M.~L. Gardel, and S.~Banerjee,
  ``Mechanosensitive junction remodeling promotes robust epithelial
  morphogenesis,'' {\em Biophysical Journal}, vol.~117, no.~9, pp.~1739--1750,
  2019.

\bibitem{cavanaugh2020rhoa}
K.~E. Cavanaugh, M.~F. Staddon, E.~Munro, S.~Banerjee, and M.~L. Gardel, ``Rhoa
  mediates epithelial cell shape changes via mechanosensitive endocytosis,''
  {\em Developmental Cell}, vol.~52, no.~2, pp.~152--166, 2020.

\bibitem{nishizawa2023two}
K.~Nishizawa, S.-Z. Lin, C.~Chard{\`e}s, J.-F. Rupprecht, and P.-F. Lenne,
  ``Two-point optical manipulation reveals mechanosensitive remodeling of
  cell--cell contacts in vivo,'' {\em Proceedings of the National Academy of
  Sciences}, vol.~120, no.~13, p.~e2212389120, 2023.

\end{thebibliography}


\begin{thebibliography}{52}%
\makeatletter
\providecommand \@ifxundefined [1]{%
 \@ifx{#1\undefined}
}%
\providecommand \@ifnum [1]{%
 \ifnum #1\expandafter \@firstoftwo
 \else \expandafter \@secondoftwo
 \fi
}%
\providecommand \@ifx [1]{%
 \ifx #1\expandafter \@firstoftwo
 \else \expandafter \@secondoftwo
 \fi
}%
\providecommand \natexlab [1]{#1}%
\providecommand \enquote  [1]{``#1''}%
\providecommand \bibnamefont  [1]{#1}%
\providecommand \bibfnamefont [1]{#1}%
\providecommand \citenamefont [1]{#1}%
\providecommand \href@noop [0]{\@secondoftwo}%
\providecommand \href [0]{\begingroup \@sanitize@url \@href}%
\providecommand \@href[1]{\@@startlink{#1}\@@href}%
\providecommand \@@href[1]{\endgroup#1\@@endlink}%
\providecommand \@sanitize@url [0]{\catcode `\\12\catcode `\$12\catcode
  `\&12\catcode `\#12\catcode `\^12\catcode `\_12\catcode `\%12\relax}%
\providecommand \@@startlink[1]{}%
\providecommand \@@endlink[0]{}%
\providecommand \url  [0]{\begingroup\@sanitize@url \@url }%
\providecommand \@url [1]{\endgroup\@href {#1}{\urlprefix }}%
\providecommand \urlprefix  [0]{URL }%
\providecommand \Eprint [0]{\href }%
\providecommand \doibase [0]{https://doi.org/}%
\providecommand \selectlanguage [0]{\@gobble}%
\providecommand \bibinfo  [0]{\@secondoftwo}%
\providecommand \bibfield  [0]{\@secondoftwo}%
\providecommand \translation [1]{[#1]}%
\providecommand \BibitemOpen [0]{}%
\providecommand \bibitemStop [0]{}%
\providecommand \bibitemNoStop [0]{.\EOS\space}%
\providecommand \EOS [0]{\spacefactor3000\relax}%
\providecommand \BibitemShut  [1]{\csname bibitem#1\endcsname}%
\let\auto@bib@innerbib\@empty
\bibitem [{\citenamefont {Mongera}\ \emph {et~al.}(2018)\citenamefont
  {Mongera}, \citenamefont {Rowghanian}, \citenamefont {Gustafson},
  \citenamefont {Shelton}, \citenamefont {Kealhofer}, \citenamefont {Carn},
  \citenamefont {Serwane}, \citenamefont {Lucio}, \citenamefont {Giammona},\
  and\ \citenamefont {Camp{\`a}s}}]{mongera2018fluid}%
  \BibitemOpen
  \bibfield  {author} {\bibinfo {author} {\bibfnamefont {A.}~\bibnamefont
  {Mongera}}, \bibinfo {author} {\bibfnamefont {P.}~\bibnamefont {Rowghanian}},
  \bibinfo {author} {\bibfnamefont {H.~J.}\ \bibnamefont {Gustafson}}, \bibinfo
  {author} {\bibfnamefont {E.}~\bibnamefont {Shelton}}, \bibinfo {author}
  {\bibfnamefont {D.~A.}\ \bibnamefont {Kealhofer}}, \bibinfo {author}
  {\bibfnamefont {E.~K.}\ \bibnamefont {Carn}}, \bibinfo {author}
  {\bibfnamefont {F.}~\bibnamefont {Serwane}}, \bibinfo {author} {\bibfnamefont
  {A.~A.}\ \bibnamefont {Lucio}}, \bibinfo {author} {\bibfnamefont
  {J.}~\bibnamefont {Giammona}},\ and\ \bibinfo {author} {\bibfnamefont
  {O.}~\bibnamefont {Camp{\`a}s}},\ }\bibfield  {title} {\bibinfo {title} {A
  fluid-to-solid jamming transition underlies vertebrate body axis
  elongation},\ }\href@noop {} {\bibfield  {journal} {\bibinfo  {journal}
  {Nature}\ }\textbf {\bibinfo {volume} {561}},\ \bibinfo {pages} {401}
  (\bibinfo {year} {2018})}\BibitemShut {NoStop}%
\bibitem [{\citenamefont {Saadaoui}\ \emph {et~al.}(2020)\citenamefont
  {Saadaoui}, \citenamefont {Rocancourt}, \citenamefont {Roussel},
  \citenamefont {Corson},\ and\ \citenamefont {Gros}}]{saadaoui2020}%
  \BibitemOpen
  \bibfield  {author} {\bibinfo {author} {\bibfnamefont {M.}~\bibnamefont
  {Saadaoui}}, \bibinfo {author} {\bibfnamefont {D.}~\bibnamefont
  {Rocancourt}}, \bibinfo {author} {\bibfnamefont {J.}~\bibnamefont {Roussel}},
  \bibinfo {author} {\bibfnamefont {F.}~\bibnamefont {Corson}},\ and\ \bibinfo
  {author} {\bibfnamefont {J.}~\bibnamefont {Gros}},\ }\bibfield  {title}
  {\bibinfo {title} {A tensile ring drives tissue flows to shape the
  gastrulating amniote embryo},\ }\href@noop {} {\bibfield  {journal} {\bibinfo
   {journal} {Science}\ }\textbf {\bibinfo {volume} {367}},\ \bibinfo {pages}
  {453} (\bibinfo {year} {2020})}\BibitemShut {NoStop}%
\bibitem [{\citenamefont {Tetley}\ \emph {et~al.}(2019)\citenamefont {Tetley},
  \citenamefont {Staddon}, \citenamefont {Heller}, \citenamefont {Hoppe},
  \citenamefont {Banerjee},\ and\ \citenamefont {Mao}}]{tetley2019tissue}%
  \BibitemOpen
  \bibfield  {author} {\bibinfo {author} {\bibfnamefont {R.~J.}\ \bibnamefont
  {Tetley}}, \bibinfo {author} {\bibfnamefont {M.~F.}\ \bibnamefont {Staddon}},
  \bibinfo {author} {\bibfnamefont {D.}~\bibnamefont {Heller}}, \bibinfo
  {author} {\bibfnamefont {A.}~\bibnamefont {Hoppe}}, \bibinfo {author}
  {\bibfnamefont {S.}~\bibnamefont {Banerjee}},\ and\ \bibinfo {author}
  {\bibfnamefont {Y.}~\bibnamefont {Mao}},\ }\bibfield  {title} {\bibinfo
  {title} {Tissue fluidity promotes epithelial wound healing},\ }\href@noop {}
  {\bibfield  {journal} {\bibinfo  {journal} {Nature Physics}\ }\textbf
  {\bibinfo {volume} {15}},\ \bibinfo {pages} {1195} (\bibinfo {year}
  {2019})}\BibitemShut {NoStop}%
\bibitem [{\citenamefont {Park}\ \emph {et~al.}(2015)\citenamefont {Park},
  \citenamefont {Kim}, \citenamefont {Bi}, \citenamefont {Mitchel},
  \citenamefont {Qazvini}, \citenamefont {Tantisira}, \citenamefont {Park},
  \citenamefont {McGill}, \citenamefont {Kim}, \citenamefont {Gweon} \emph
  {et~al.}}]{park2015}%
  \BibitemOpen
  \bibfield  {author} {\bibinfo {author} {\bibfnamefont {J.-A.}\ \bibnamefont
  {Park}}, \bibinfo {author} {\bibfnamefont {J.~H.}\ \bibnamefont {Kim}},
  \bibinfo {author} {\bibfnamefont {D.}~\bibnamefont {Bi}}, \bibinfo {author}
  {\bibfnamefont {J.~A.}\ \bibnamefont {Mitchel}}, \bibinfo {author}
  {\bibfnamefont {N.~T.}\ \bibnamefont {Qazvini}}, \bibinfo {author}
  {\bibfnamefont {K.}~\bibnamefont {Tantisira}}, \bibinfo {author}
  {\bibfnamefont {C.~Y.}\ \bibnamefont {Park}}, \bibinfo {author}
  {\bibfnamefont {M.}~\bibnamefont {McGill}}, \bibinfo {author} {\bibfnamefont
  {S.-H.}\ \bibnamefont {Kim}}, \bibinfo {author} {\bibfnamefont
  {B.}~\bibnamefont {Gweon}}, \emph {et~al.},\ }\bibfield  {title} {\bibinfo
  {title} {Unjamming and cell shape in the asthmatic airway epithelium},\
  }\href@noop {} {\bibfield  {journal} {\bibinfo  {journal} {Nature Materials}\
  }\textbf {\bibinfo {volume} {14}},\ \bibinfo {pages} {1040} (\bibinfo {year}
  {2015})}\BibitemShut {NoStop}%
\bibitem [{\citenamefont {Atia}\ \emph {et~al.}(2018)\citenamefont {Atia},
  \citenamefont {Bi}, \citenamefont {Sharma}, \citenamefont {Mitchel},
  \citenamefont {Gweon}, \citenamefont {A~Koehler}, \citenamefont {DeCamp},
  \citenamefont {Lan}, \citenamefont {Kim}, \citenamefont {Hirsch} \emph
  {et~al.}}]{atia2018}%
  \BibitemOpen
  \bibfield  {author} {\bibinfo {author} {\bibfnamefont {L.}~\bibnamefont
  {Atia}}, \bibinfo {author} {\bibfnamefont {D.}~\bibnamefont {Bi}}, \bibinfo
  {author} {\bibfnamefont {Y.}~\bibnamefont {Sharma}}, \bibinfo {author}
  {\bibfnamefont {J.~A.}\ \bibnamefont {Mitchel}}, \bibinfo {author}
  {\bibfnamefont {B.}~\bibnamefont {Gweon}}, \bibinfo {author} {\bibfnamefont
  {S.}~\bibnamefont {A~Koehler}}, \bibinfo {author} {\bibfnamefont {S.~J.}\
  \bibnamefont {DeCamp}}, \bibinfo {author} {\bibfnamefont {B.}~\bibnamefont
  {Lan}}, \bibinfo {author} {\bibfnamefont {J.~H.}\ \bibnamefont {Kim}},
  \bibinfo {author} {\bibfnamefont {R.}~\bibnamefont {Hirsch}}, \emph
  {et~al.},\ }\bibfield  {title} {\bibinfo {title} {Geometric constraints
  during epithelial jamming},\ }\href@noop {} {\bibfield  {journal} {\bibinfo
  {journal} {Nature Physics}\ }\textbf {\bibinfo {volume} {14}},\ \bibinfo
  {pages} {613} (\bibinfo {year} {2018})}\BibitemShut {NoStop}%
\bibitem [{\citenamefont {Barriga}\ \emph {et~al.}(2018)\citenamefont
  {Barriga}, \citenamefont {Franze}, \citenamefont {Charras},\ and\
  \citenamefont {Mayor}}]{barriga2018}%
  \BibitemOpen
  \bibfield  {author} {\bibinfo {author} {\bibfnamefont {E.~H.}\ \bibnamefont
  {Barriga}}, \bibinfo {author} {\bibfnamefont {K.}~\bibnamefont {Franze}},
  \bibinfo {author} {\bibfnamefont {G.}~\bibnamefont {Charras}},\ and\ \bibinfo
  {author} {\bibfnamefont {R.}~\bibnamefont {Mayor}},\ }\bibfield  {title}
  {\bibinfo {title} {Tissue stiffening coordinates morphogenesis by triggering
  collective cell migration in vivo},\ }\href@noop {} {\bibfield  {journal}
  {\bibinfo  {journal} {Nature}\ }\textbf {\bibinfo {volume} {554}},\ \bibinfo
  {pages} {523} (\bibinfo {year} {2018})}\BibitemShut {NoStop}%
\bibitem [{\citenamefont {Ranft}\ \emph {et~al.}(2010)\citenamefont {Ranft},
  \citenamefont {Basan}, \citenamefont {Elgeti}, \citenamefont {Joanny},
  \citenamefont {Prost},\ and\ \citenamefont {J{\"u}licher}}]{ranft2010}%
  \BibitemOpen
  \bibfield  {author} {\bibinfo {author} {\bibfnamefont {J.}~\bibnamefont
  {Ranft}}, \bibinfo {author} {\bibfnamefont {M.}~\bibnamefont {Basan}},
  \bibinfo {author} {\bibfnamefont {J.}~\bibnamefont {Elgeti}}, \bibinfo
  {author} {\bibfnamefont {J.-F.}\ \bibnamefont {Joanny}}, \bibinfo {author}
  {\bibfnamefont {J.}~\bibnamefont {Prost}},\ and\ \bibinfo {author}
  {\bibfnamefont {F.}~\bibnamefont {J{\"u}licher}},\ }\bibfield  {title}
  {\bibinfo {title} {Fluidization of tissues by cell division and apoptosis},\
  }\href@noop {} {\bibfield  {journal} {\bibinfo  {journal} {Proceedings of the
  National Academy of Sciences}\ }\textbf {\bibinfo {volume} {107}},\ \bibinfo
  {pages} {20863} (\bibinfo {year} {2010})}\BibitemShut {NoStop}%
\bibitem [{\citenamefont {Tetley}\ and\ \citenamefont
  {Mao}(2018)}]{tetley2018}%
  \BibitemOpen
  \bibfield  {author} {\bibinfo {author} {\bibfnamefont {R.~J.}\ \bibnamefont
  {Tetley}}\ and\ \bibinfo {author} {\bibfnamefont {Y.}~\bibnamefont {Mao}},\
  }\bibfield  {title} {\bibinfo {title} {The same but different: cell
  intercalation as a driver of tissue deformation and fluidity},\ }\href@noop
  {} {\bibfield  {journal} {\bibinfo  {journal} {Philosophical Transactions of
  the Royal Society B: Biological Sciences}\ }\textbf {\bibinfo {volume}
  {373}},\ \bibinfo {pages} {20170328} (\bibinfo {year} {2018})}\BibitemShut
  {NoStop}%
\bibitem [{\citenamefont {Lemke}\ and\ \citenamefont
  {Nelson}(2021)}]{lemke2021dynamic}%
  \BibitemOpen
  \bibfield  {author} {\bibinfo {author} {\bibfnamefont {S.~B.}\ \bibnamefont
  {Lemke}}\ and\ \bibinfo {author} {\bibfnamefont {C.~M.}\ \bibnamefont
  {Nelson}},\ }\bibfield  {title} {\bibinfo {title} {Dynamic changes in
  epithelial cell packing during tissue morphogenesis},\ }\href@noop {}
  {\bibfield  {journal} {\bibinfo  {journal} {Current Biology}\ }\textbf
  {\bibinfo {volume} {31}},\ \bibinfo {pages} {R1098} (\bibinfo {year}
  {2021})}\BibitemShut {NoStop}%
\bibitem [{\citenamefont {Classen}\ \emph {et~al.}(2005)\citenamefont
  {Classen}, \citenamefont {Anderson}, \citenamefont {Marois},\ and\
  \citenamefont {Eaton}}]{classen2005hexagonal}%
  \BibitemOpen
  \bibfield  {author} {\bibinfo {author} {\bibfnamefont {A.-K.}\ \bibnamefont
  {Classen}}, \bibinfo {author} {\bibfnamefont {K.~I.}\ \bibnamefont
  {Anderson}}, \bibinfo {author} {\bibfnamefont {E.}~\bibnamefont {Marois}},\
  and\ \bibinfo {author} {\bibfnamefont {S.}~\bibnamefont {Eaton}},\ }\bibfield
   {title} {\bibinfo {title} {Hexagonal packing of drosophila wing epithelial
  cells by the planar cell polarity pathway},\ }\href@noop {} {\bibfield
  {journal} {\bibinfo  {journal} {Developmental cell}\ }\textbf {\bibinfo
  {volume} {9}},\ \bibinfo {pages} {805} (\bibinfo {year} {2005})}\BibitemShut
  {NoStop}%
\bibitem [{\citenamefont {Blankenship}\ \emph {et~al.}(2006)\citenamefont
  {Blankenship}, \citenamefont {Backovic}, \citenamefont {Sanny}, \citenamefont
  {Weitz},\ and\ \citenamefont {Zallen}}]{blankenship2006multicellular}%
  \BibitemOpen
  \bibfield  {author} {\bibinfo {author} {\bibfnamefont {J.~T.}\ \bibnamefont
  {Blankenship}}, \bibinfo {author} {\bibfnamefont {S.~T.}\ \bibnamefont
  {Backovic}}, \bibinfo {author} {\bibfnamefont {J.~S.}\ \bibnamefont {Sanny}},
  \bibinfo {author} {\bibfnamefont {O.}~\bibnamefont {Weitz}},\ and\ \bibinfo
  {author} {\bibfnamefont {J.~A.}\ \bibnamefont {Zallen}},\ }\bibfield  {title}
  {\bibinfo {title} {Multicellular rosette formation links planar cell polarity
  to tissue morphogenesis},\ }\href@noop {} {\bibfield  {journal} {\bibinfo
  {journal} {Developmental Cell}\ }\textbf {\bibinfo {volume} {11}},\ \bibinfo
  {pages} {459} (\bibinfo {year} {2006})}\BibitemShut {NoStop}%
\bibitem [{\citenamefont {Jessica}\ and\ \citenamefont
  {Fernandez-Gonzalez}(2016)}]{jessica2016local}%
  \BibitemOpen
  \bibfield  {author} {\bibinfo {author} {\bibfnamefont {C.~Y.}\ \bibnamefont
  {Jessica}}\ and\ \bibinfo {author} {\bibfnamefont {R.}~\bibnamefont
  {Fernandez-Gonzalez}},\ }\bibfield  {title} {\bibinfo {title} {Local
  mechanical forces promote polarized junctional assembly and axis elongation
  in drosophila},\ }\href@noop {} {\bibfield  {journal} {\bibinfo  {journal}
  {eLife}\ }\textbf {\bibinfo {volume} {5}},\ \bibinfo {pages} {e10757}
  (\bibinfo {year} {2016})}\BibitemShut {NoStop}%
\bibitem [{\citenamefont {Finegan}\ \emph {et~al.}(2019)\citenamefont
  {Finegan}, \citenamefont {Hervieux}, \citenamefont {Nestor-Bergmann},
  \citenamefont {Fletcher}, \citenamefont {Blanchard},\ and\ \citenamefont
  {Sanson}}]{finegan2019tricellular}%
  \BibitemOpen
  \bibfield  {author} {\bibinfo {author} {\bibfnamefont {T.~M.}\ \bibnamefont
  {Finegan}}, \bibinfo {author} {\bibfnamefont {N.}~\bibnamefont {Hervieux}},
  \bibinfo {author} {\bibfnamefont {A.}~\bibnamefont {Nestor-Bergmann}},
  \bibinfo {author} {\bibfnamefont {A.~G.}\ \bibnamefont {Fletcher}}, \bibinfo
  {author} {\bibfnamefont {G.~B.}\ \bibnamefont {Blanchard}},\ and\ \bibinfo
  {author} {\bibfnamefont {B.}~\bibnamefont {Sanson}},\ }\bibfield  {title}
  {\bibinfo {title} {The tricellular vertex-specific adhesion molecule sidekick
  facilitates polarised cell intercalation during drosophila axis extension},\
  }\href@noop {} {\bibfield  {journal} {\bibinfo  {journal} {PLoS Biology}\
  }\textbf {\bibinfo {volume} {17}},\ \bibinfo {pages} {e3000522} (\bibinfo
  {year} {2019})}\BibitemShut {NoStop}%
\bibitem [{\citenamefont {Fletcher}\ \emph {et~al.}(2014)\citenamefont
  {Fletcher}, \citenamefont {Osterfield}, \citenamefont {Baker},\ and\
  \citenamefont {Shvartsman}}]{fletcher2014}%
  \BibitemOpen
  \bibfield  {author} {\bibinfo {author} {\bibfnamefont {A.~G.}\ \bibnamefont
  {Fletcher}}, \bibinfo {author} {\bibfnamefont {M.}~\bibnamefont
  {Osterfield}}, \bibinfo {author} {\bibfnamefont {R.~E.}\ \bibnamefont
  {Baker}},\ and\ \bibinfo {author} {\bibfnamefont {S.~Y.}\ \bibnamefont
  {Shvartsman}},\ }\bibfield  {title} {\bibinfo {title} {Vertex models of
  epithelial morphogenesis},\ }\href@noop {} {\bibfield  {journal} {\bibinfo
  {journal} {Biophysical journal}\ }\textbf {\bibinfo {volume} {106}},\
  \bibinfo {pages} {2291} (\bibinfo {year} {2014})}\BibitemShut {NoStop}%
\bibitem [{\citenamefont {Alt}\ \emph {et~al.}(2017)\citenamefont {Alt},
  \citenamefont {Ganguly},\ and\ \citenamefont {Salbreux}}]{alt2017}%
  \BibitemOpen
  \bibfield  {author} {\bibinfo {author} {\bibfnamefont {S.}~\bibnamefont
  {Alt}}, \bibinfo {author} {\bibfnamefont {P.}~\bibnamefont {Ganguly}},\ and\
  \bibinfo {author} {\bibfnamefont {G.}~\bibnamefont {Salbreux}},\ }\bibfield
  {title} {\bibinfo {title} {Vertex models: from cell mechanics to tissue
  morphogenesis},\ }\href@noop {} {\bibfield  {journal} {\bibinfo  {journal}
  {Philosophical Transactions of the Royal Society B: Biological Sciences}\
  }\textbf {\bibinfo {volume} {372}},\ \bibinfo {pages} {20150520} (\bibinfo
  {year} {2017})}\BibitemShut {NoStop}%
\bibitem [{\citenamefont {Spencer}\ \emph {et~al.}(2017)\citenamefont
  {Spencer}, \citenamefont {Jabeen},\ and\ \citenamefont
  {Lubensky}}]{spencer2017vertex}%
  \BibitemOpen
  \bibfield  {author} {\bibinfo {author} {\bibfnamefont {M.~A.}\ \bibnamefont
  {Spencer}}, \bibinfo {author} {\bibfnamefont {Z.}~\bibnamefont {Jabeen}},\
  and\ \bibinfo {author} {\bibfnamefont {D.~K.}\ \bibnamefont {Lubensky}},\
  }\bibfield  {title} {\bibinfo {title} {Vertex stability and topological
  transitions in vertex models of foams and epithelia},\ }\href@noop {}
  {\bibfield  {journal} {\bibinfo  {journal} {The European Physical Journal E}\
  }\textbf {\bibinfo {volume} {40}},\ \bibinfo {pages} {1} (\bibinfo {year}
  {2017})}\BibitemShut {NoStop}%
\bibitem [{\citenamefont {Curran}\ \emph {et~al.}(2017)\citenamefont {Curran},
  \citenamefont {Strandkvist}, \citenamefont {Bathmann}, \citenamefont
  {de~Gennes}, \citenamefont {Kabla}, \citenamefont {Salbreux},\ and\
  \citenamefont {Baum}}]{curran2017myosin}%
  \BibitemOpen
  \bibfield  {author} {\bibinfo {author} {\bibfnamefont {S.}~\bibnamefont
  {Curran}}, \bibinfo {author} {\bibfnamefont {C.}~\bibnamefont {Strandkvist}},
  \bibinfo {author} {\bibfnamefont {J.}~\bibnamefont {Bathmann}}, \bibinfo
  {author} {\bibfnamefont {M.}~\bibnamefont {de~Gennes}}, \bibinfo {author}
  {\bibfnamefont {A.}~\bibnamefont {Kabla}}, \bibinfo {author} {\bibfnamefont
  {G.}~\bibnamefont {Salbreux}},\ and\ \bibinfo {author} {\bibfnamefont
  {B.}~\bibnamefont {Baum}},\ }\bibfield  {title} {\bibinfo {title} {Myosin ii
  controls junction fluctuations to guide epithelial tissue ordering},\
  }\href@noop {} {\bibfield  {journal} {\bibinfo  {journal} {Developmental
  Cell}\ }\textbf {\bibinfo {volume} {43}},\ \bibinfo {pages} {480} (\bibinfo
  {year} {2017})}\BibitemShut {NoStop}%
\bibitem [{\citenamefont {Comelles}\ \emph {et~al.}(2021)\citenamefont
  {Comelles}, \citenamefont {Soumya}, \citenamefont {Lu}, \citenamefont
  {Le~Maout}, \citenamefont {Anvitha}, \citenamefont {Salbreux}, \citenamefont
  {J{\"u}licher}, \citenamefont {Inamdar},\ and\ \citenamefont
  {Riveline}}]{comelles2021epithelial}%
  \BibitemOpen
  \bibfield  {author} {\bibinfo {author} {\bibfnamefont {J.}~\bibnamefont
  {Comelles}}, \bibinfo {author} {\bibfnamefont {S.}~\bibnamefont {Soumya}},
  \bibinfo {author} {\bibfnamefont {L.}~\bibnamefont {Lu}}, \bibinfo {author}
  {\bibfnamefont {E.}~\bibnamefont {Le~Maout}}, \bibinfo {author}
  {\bibfnamefont {S.}~\bibnamefont {Anvitha}}, \bibinfo {author} {\bibfnamefont
  {G.}~\bibnamefont {Salbreux}}, \bibinfo {author} {\bibfnamefont
  {F.}~\bibnamefont {J{\"u}licher}}, \bibinfo {author} {\bibfnamefont {M.~M.}\
  \bibnamefont {Inamdar}},\ and\ \bibinfo {author} {\bibfnamefont
  {D.}~\bibnamefont {Riveline}},\ }\bibfield  {title} {\bibinfo {title}
  {Epithelial colonies in vitro elongate through collective effects},\
  }\href@noop {} {\bibfield  {journal} {\bibinfo  {journal} {eLife}\ }\textbf
  {\bibinfo {volume} {10}} (\bibinfo {year} {2021})}\BibitemShut {NoStop}%
\bibitem [{\citenamefont {Staddon}\ \emph {et~al.}(2018)\citenamefont
  {Staddon}, \citenamefont {Bi}, \citenamefont {Tabatabai}, \citenamefont
  {Ajeti}, \citenamefont {Murrell},\ and\ \citenamefont
  {Banerjee}}]{staddon2018}%
  \BibitemOpen
  \bibfield  {author} {\bibinfo {author} {\bibfnamefont {M.~F.}\ \bibnamefont
  {Staddon}}, \bibinfo {author} {\bibfnamefont {D.}~\bibnamefont {Bi}},
  \bibinfo {author} {\bibfnamefont {A.~P.}\ \bibnamefont {Tabatabai}}, \bibinfo
  {author} {\bibfnamefont {V.}~\bibnamefont {Ajeti}}, \bibinfo {author}
  {\bibfnamefont {M.~P.}\ \bibnamefont {Murrell}},\ and\ \bibinfo {author}
  {\bibfnamefont {S.}~\bibnamefont {Banerjee}},\ }\bibfield  {title} {\bibinfo
  {title} {Cooperation of dual modes of cell motility promotes epithelial
  stress relaxation to accelerate wound healing},\ }\href@noop {} {\bibfield
  {journal} {\bibinfo  {journal} {PLoS Computational Biology}\ }\textbf
  {\bibinfo {volume} {14}},\ \bibinfo {pages} {e1006502} (\bibinfo {year}
  {2018})}\BibitemShut {NoStop}%
\bibitem [{\citenamefont {Yamamoto}\ \emph {et~al.}(2022)\citenamefont
  {Yamamoto}, \citenamefont {Sussman}, \citenamefont {Shibata},\ and\
  \citenamefont {Manning}}]{yamamoto2022non}%
  \BibitemOpen
  \bibfield  {author} {\bibinfo {author} {\bibfnamefont {T.}~\bibnamefont
  {Yamamoto}}, \bibinfo {author} {\bibfnamefont {D.~M.}\ \bibnamefont
  {Sussman}}, \bibinfo {author} {\bibfnamefont {T.}~\bibnamefont {Shibata}},\
  and\ \bibinfo {author} {\bibfnamefont {M.~L.}\ \bibnamefont {Manning}},\
  }\bibfield  {title} {\bibinfo {title} {Non-monotonic fluidization generated
  by fluctuating edge tensions in confluent tissues},\ }\href@noop {}
  {\bibfield  {journal} {\bibinfo  {journal} {Soft Matter}\ }\textbf {\bibinfo
  {volume} {18}},\ \bibinfo {pages} {2168} (\bibinfo {year}
  {2022})}\BibitemShut {NoStop}%
\bibitem [{\citenamefont {Farhadifar}\ \emph {et~al.}(2007)\citenamefont
  {Farhadifar}, \citenamefont {R{\"o}per}, \citenamefont {Aigouy},
  \citenamefont {Eaton},\ and\ \citenamefont
  {J{\"u}licher}}]{farhadifar2007influence}%
  \BibitemOpen
  \bibfield  {author} {\bibinfo {author} {\bibfnamefont {R.}~\bibnamefont
  {Farhadifar}}, \bibinfo {author} {\bibfnamefont {J.-C.}\ \bibnamefont
  {R{\"o}per}}, \bibinfo {author} {\bibfnamefont {B.}~\bibnamefont {Aigouy}},
  \bibinfo {author} {\bibfnamefont {S.}~\bibnamefont {Eaton}},\ and\ \bibinfo
  {author} {\bibfnamefont {F.}~\bibnamefont {J{\"u}licher}},\ }\bibfield
  {title} {\bibinfo {title} {The influence of cell mechanics, cell-cell
  interactions, and proliferation on epithelial packing},\ }\href@noop {}
  {\bibfield  {journal} {\bibinfo  {journal} {Current Biology}\ }\textbf
  {\bibinfo {volume} {17}},\ \bibinfo {pages} {2095} (\bibinfo {year}
  {2007})}\BibitemShut {NoStop}%
\bibitem [{\citenamefont {Farhadifar}(2009)}]{farhadifar2009dynamics}%
  \BibitemOpen
  \bibfield  {author} {\bibinfo {author} {\bibfnamefont {R.}~\bibnamefont
  {Farhadifar}},\ }\emph {\bibinfo {title} {Dynamics of cell packing and polar
  order in developing epithelia}},\ \href@noop {} {Ph.D. thesis},\ \bibinfo
  {school} {Dresden, Techn. Univ., Diss., 2009} (\bibinfo {year}
  {2009})\BibitemShut {NoStop}%
\bibitem [{\citenamefont {Yan}\ and\ \citenamefont
  {Bi}(2019)}]{yan2019multicellular}%
  \BibitemOpen
  \bibfield  {author} {\bibinfo {author} {\bibfnamefont {L.}~\bibnamefont
  {Yan}}\ and\ \bibinfo {author} {\bibfnamefont {D.}~\bibnamefont {Bi}},\
  }\bibfield  {title} {\bibinfo {title} {Multicellular rosettes drive
  fluid-solid transition in epithelial tissues},\ }\href@noop {} {\bibfield
  {journal} {\bibinfo  {journal} {Physical Review X}\ }\textbf {\bibinfo
  {volume} {9}},\ \bibinfo {pages} {011029} (\bibinfo {year}
  {2019})}\BibitemShut {NoStop}%
\bibitem [{\citenamefont {Das}\ \emph {et~al.}(2021)\citenamefont {Das},
  \citenamefont {Sastry},\ and\ \citenamefont {Bi}}]{das2021controlled}%
  \BibitemOpen
  \bibfield  {author} {\bibinfo {author} {\bibfnamefont {A.}~\bibnamefont
  {Das}}, \bibinfo {author} {\bibfnamefont {S.}~\bibnamefont {Sastry}},\ and\
  \bibinfo {author} {\bibfnamefont {D.}~\bibnamefont {Bi}},\ }\bibfield
  {title} {\bibinfo {title} {Controlled neighbor exchanges drive glassy
  behavior, intermittency, and cell streaming in epithelial tissues},\
  }\href@noop {} {\bibfield  {journal} {\bibinfo  {journal} {Physical Review
  X}\ }\textbf {\bibinfo {volume} {11}},\ \bibinfo {pages} {041037} (\bibinfo
  {year} {2021})}\BibitemShut {NoStop}%
\bibitem [{\citenamefont {Erdemci-Tandogan}\ and\ \citenamefont
  {Manning}(2021)}]{erdemci2021effect}%
  \BibitemOpen
  \bibfield  {author} {\bibinfo {author} {\bibfnamefont {G.}~\bibnamefont
  {Erdemci-Tandogan}}\ and\ \bibinfo {author} {\bibfnamefont {M.~L.}\
  \bibnamefont {Manning}},\ }\bibfield  {title} {\bibinfo {title} {Effect of
  cellular rearrangement time delays on the rheology of vertex models for
  confluent tissues},\ }\href@noop {} {\bibfield  {journal} {\bibinfo
  {journal} {PLoS Computational Biology}\ }\textbf {\bibinfo {volume} {17}},\
  \bibinfo {pages} {e1009049} (\bibinfo {year} {2021})}\BibitemShut {NoStop}%
\bibitem [{\citenamefont {Nagai}\ and\ \citenamefont
  {Honda}(2001)}]{nagai2001dynamic}%
  \BibitemOpen
  \bibfield  {author} {\bibinfo {author} {\bibfnamefont {T.}~\bibnamefont
  {Nagai}}\ and\ \bibinfo {author} {\bibfnamefont {H.}~\bibnamefont {Honda}},\
  }\bibfield  {title} {\bibinfo {title} {A dynamic cell model for the formation
  of epithelial tissues},\ }\href@noop {} {\bibfield  {journal} {\bibinfo
  {journal} {Philosophical Magazine B}\ }\textbf {\bibinfo {volume} {81}},\
  \bibinfo {pages} {699} (\bibinfo {year} {2001})}\BibitemShut {NoStop}%
\bibitem [{\citenamefont {Staple}\ \emph {et~al.}(2010)\citenamefont {Staple},
  \citenamefont {Farhadifar}, \citenamefont {R{\"o}per}, \citenamefont
  {Aigouy}, \citenamefont {Eaton},\ and\ \citenamefont
  {J{\"u}licher}}]{staple2010mechanics}%
  \BibitemOpen
  \bibfield  {author} {\bibinfo {author} {\bibfnamefont {D.~B.}\ \bibnamefont
  {Staple}}, \bibinfo {author} {\bibfnamefont {R.}~\bibnamefont {Farhadifar}},
  \bibinfo {author} {\bibfnamefont {J.-C.}\ \bibnamefont {R{\"o}per}}, \bibinfo
  {author} {\bibfnamefont {B.}~\bibnamefont {Aigouy}}, \bibinfo {author}
  {\bibfnamefont {S.}~\bibnamefont {Eaton}},\ and\ \bibinfo {author}
  {\bibfnamefont {F.}~\bibnamefont {J{\"u}licher}},\ }\bibfield  {title}
  {\bibinfo {title} {Mechanics and remodelling of cell packings in epithelia},\
  }\href@noop {} {\bibfield  {journal} {\bibinfo  {journal} {The European
  Physical Journal E}\ }\textbf {\bibinfo {volume} {33}},\ \bibinfo {pages}
  {117} (\bibinfo {year} {2010})}\BibitemShut {NoStop}%
\bibitem [{\citenamefont {Staddon}\ \emph {et~al.}(2019)\citenamefont
  {Staddon}, \citenamefont {Cavanaugh}, \citenamefont {Munro}, \citenamefont
  {Gardel},\ and\ \citenamefont {Banerjee}}]{staddon2019mechanosensitive}%
  \BibitemOpen
  \bibfield  {author} {\bibinfo {author} {\bibfnamefont {M.~F.}\ \bibnamefont
  {Staddon}}, \bibinfo {author} {\bibfnamefont {K.~E.}\ \bibnamefont
  {Cavanaugh}}, \bibinfo {author} {\bibfnamefont {E.~M.}\ \bibnamefont
  {Munro}}, \bibinfo {author} {\bibfnamefont {M.~L.}\ \bibnamefont {Gardel}},\
  and\ \bibinfo {author} {\bibfnamefont {S.}~\bibnamefont {Banerjee}},\
  }\bibfield  {title} {\bibinfo {title} {Mechanosensitive junction remodeling
  promotes robust epithelial morphogenesis},\ }\href@noop {} {\bibfield
  {journal} {\bibinfo  {journal} {Biophysical Journal}\ }\textbf {\bibinfo
  {volume} {117}},\ \bibinfo {pages} {1739} (\bibinfo {year}
  {2019})}\BibitemShut {NoStop}%
\bibitem [{\citenamefont {Krajnc}\ \emph {et~al.}(2021)\citenamefont {Krajnc},
  \citenamefont {Stern},\ and\ \citenamefont {Zankoc}}]{krajnc2021active}%
  \BibitemOpen
  \bibfield  {author} {\bibinfo {author} {\bibfnamefont {M.}~\bibnamefont
  {Krajnc}}, \bibinfo {author} {\bibfnamefont {T.}~\bibnamefont {Stern}},\ and\
  \bibinfo {author} {\bibfnamefont {C.}~\bibnamefont {Zankoc}},\ }\bibfield
  {title} {\bibinfo {title} {Active instability and nonlinear dynamics of
  cell-cell junctions},\ }\href@noop {} {\bibfield  {journal} {\bibinfo
  {journal} {Physical Review Letters}\ }\textbf {\bibinfo {volume} {127}},\
  \bibinfo {pages} {198103} (\bibinfo {year} {2021})}\BibitemShut {NoStop}%
\bibitem [{\citenamefont {Gustafson}\ \emph
  {et~al.}(2022{\natexlab{a}})\citenamefont {Gustafson}, \citenamefont
  {Claussen}, \citenamefont {De~Renzis},\ and\ \citenamefont
  {Streichan}}]{gustafson2022patterned}%
  \BibitemOpen
  \bibfield  {author} {\bibinfo {author} {\bibfnamefont {H.~J.}\ \bibnamefont
  {Gustafson}}, \bibinfo {author} {\bibfnamefont {N.}~\bibnamefont {Claussen}},
  \bibinfo {author} {\bibfnamefont {S.}~\bibnamefont {De~Renzis}},\ and\
  \bibinfo {author} {\bibfnamefont {S.~J.}\ \bibnamefont {Streichan}},\
  }\bibfield  {title} {\bibinfo {title} {Patterned mechanical feedback
  establishes a global myosin gradient},\ }\href@noop {} {\bibfield  {journal}
  {\bibinfo  {journal} {Nature Communications}\ }\textbf {\bibinfo {volume}
  {13}},\ \bibinfo {pages} {7050} (\bibinfo {year}
  {2022}{\natexlab{a}})}\BibitemShut {NoStop}%
\bibitem [{\citenamefont {Cavanaugh}\ \emph
  {et~al.}(2020{\natexlab{a}})\citenamefont {Cavanaugh}, \citenamefont
  {Staddon}, \citenamefont {Munro}, \citenamefont {Banerjee},\ and\
  \citenamefont {Gardel}}]{cavanaugh2020rhoa}%
  \BibitemOpen
  \bibfield  {author} {\bibinfo {author} {\bibfnamefont {K.~E.}\ \bibnamefont
  {Cavanaugh}}, \bibinfo {author} {\bibfnamefont {M.~F.}\ \bibnamefont
  {Staddon}}, \bibinfo {author} {\bibfnamefont {E.}~\bibnamefont {Munro}},
  \bibinfo {author} {\bibfnamefont {S.}~\bibnamefont {Banerjee}},\ and\
  \bibinfo {author} {\bibfnamefont {M.~L.}\ \bibnamefont {Gardel}},\ }\bibfield
   {title} {\bibinfo {title} {Rhoa mediates epithelial cell shape changes via
  mechanosensitive endocytosis},\ }\href@noop {} {\bibfield  {journal}
  {\bibinfo  {journal} {Developmental Cell}\ }\textbf {\bibinfo {volume}
  {52}},\ \bibinfo {pages} {152} (\bibinfo {year}
  {2020}{\natexlab{a}})}\BibitemShut {NoStop}%
\bibitem [{\citenamefont {Nishizawa}\ \emph {et~al.}(2023)\citenamefont
  {Nishizawa}, \citenamefont {Lin}, \citenamefont {Chard{\`e}s}, \citenamefont
  {Rupprecht},\ and\ \citenamefont {Lenne}}]{nishizawa2022}%
  \BibitemOpen
  \bibfield  {author} {\bibinfo {author} {\bibfnamefont {K.}~\bibnamefont
  {Nishizawa}}, \bibinfo {author} {\bibfnamefont {S.-Z.}\ \bibnamefont {Lin}},
  \bibinfo {author} {\bibfnamefont {C.}~\bibnamefont {Chard{\`e}s}}, \bibinfo
  {author} {\bibfnamefont {J.-F.}\ \bibnamefont {Rupprecht}},\ and\ \bibinfo
  {author} {\bibfnamefont {P.-F.}\ \bibnamefont {Lenne}},\ }\bibfield  {title}
  {\bibinfo {title} {Two-point optical manipulation reveals mechanosensitive
  remodeling of cell--cell contacts in vivo},\ }\href@noop {} {\bibfield
  {journal} {\bibinfo  {journal} {Proceedings of the National Academy of
  Sciences}\ }\textbf {\bibinfo {volume} {120}},\ \bibinfo {pages}
  {e2212389120} (\bibinfo {year} {2023})}\BibitemShut {NoStop}%
\bibitem [{\citenamefont {Iyer}\ \emph {et~al.}(2019)\citenamefont {Iyer},
  \citenamefont {Piscitello-G{\'o}mez}, \citenamefont {Paijmans}, \citenamefont
  {J{\"u}licher},\ and\ \citenamefont {Eaton}}]{iyer2019}%
  \BibitemOpen
  \bibfield  {author} {\bibinfo {author} {\bibfnamefont {K.~V.}\ \bibnamefont
  {Iyer}}, \bibinfo {author} {\bibfnamefont {R.}~\bibnamefont
  {Piscitello-G{\'o}mez}}, \bibinfo {author} {\bibfnamefont {J.}~\bibnamefont
  {Paijmans}}, \bibinfo {author} {\bibfnamefont {F.}~\bibnamefont
  {J{\"u}licher}},\ and\ \bibinfo {author} {\bibfnamefont {S.}~\bibnamefont
  {Eaton}},\ }\bibfield  {title} {\bibinfo {title} {Epithelial viscoelasticity
  is regulated by mechanosensitive e-cadherin turnover},\ }\href@noop {}
  {\bibfield  {journal} {\bibinfo  {journal} {Current Biology}\ }\textbf
  {\bibinfo {volume} {29}},\ \bibinfo {pages} {578} (\bibinfo {year}
  {2019})}\BibitemShut {NoStop}%
\bibitem [{\citenamefont {Cavanaugh}\ \emph {et~al.}(2022)\citenamefont
  {Cavanaugh}, \citenamefont {Staddon}, \citenamefont {Chmiel}, \citenamefont
  {Harmon}, \citenamefont {Budnar}, \citenamefont {Banerjee}, \citenamefont
  {Gardel} \emph {et~al.}}]{cavanaugh2022}%
  \BibitemOpen
  \bibfield  {author} {\bibinfo {author} {\bibfnamefont {K.~E.}\ \bibnamefont
  {Cavanaugh}}, \bibinfo {author} {\bibfnamefont {M.~F.}\ \bibnamefont
  {Staddon}}, \bibinfo {author} {\bibfnamefont {T.~A.}\ \bibnamefont {Chmiel}},
  \bibinfo {author} {\bibfnamefont {R.}~\bibnamefont {Harmon}}, \bibinfo
  {author} {\bibfnamefont {S.}~\bibnamefont {Budnar}}, \bibinfo {author}
  {\bibfnamefont {S.}~\bibnamefont {Banerjee}}, \bibinfo {author}
  {\bibfnamefont {M.~L.}\ \bibnamefont {Gardel}}, \emph {et~al.},\ }\bibfield
  {title} {\bibinfo {title} {Force-dependent intercellular adhesion
  strengthening underlies asymmetric adherens junction contraction},\
  }\href@noop {} {\bibfield  {journal} {\bibinfo  {journal} {Current Biology}\
  }\textbf {\bibinfo {volume} {32}},\ \bibinfo {pages} {1986} (\bibinfo {year}
  {2022})}\BibitemShut {NoStop}%
\bibitem [{\citenamefont {Khalilgharibi}\ \emph {et~al.}(2019)\citenamefont
  {Khalilgharibi}, \citenamefont {Fouchard}, \citenamefont {Asadipour},
  \citenamefont {Barrientos}, \citenamefont {Duda}, \citenamefont {Bonfanti},
  \citenamefont {Yonis}, \citenamefont {Harris}, \citenamefont {Mosaffa},
  \citenamefont {Fujita} \emph {et~al.}}]{khalilgharibi2019stress}%
  \BibitemOpen
  \bibfield  {author} {\bibinfo {author} {\bibfnamefont {N.}~\bibnamefont
  {Khalilgharibi}}, \bibinfo {author} {\bibfnamefont {J.}~\bibnamefont
  {Fouchard}}, \bibinfo {author} {\bibfnamefont {N.}~\bibnamefont {Asadipour}},
  \bibinfo {author} {\bibfnamefont {R.}~\bibnamefont {Barrientos}}, \bibinfo
  {author} {\bibfnamefont {M.}~\bibnamefont {Duda}}, \bibinfo {author}
  {\bibfnamefont {A.}~\bibnamefont {Bonfanti}}, \bibinfo {author}
  {\bibfnamefont {A.}~\bibnamefont {Yonis}}, \bibinfo {author} {\bibfnamefont
  {A.}~\bibnamefont {Harris}}, \bibinfo {author} {\bibfnamefont
  {P.}~\bibnamefont {Mosaffa}}, \bibinfo {author} {\bibfnamefont
  {Y.}~\bibnamefont {Fujita}}, \emph {et~al.},\ }\bibfield  {title} {\bibinfo
  {title} {Stress relaxation in epithelial monolayers is controlled by the
  actomyosin cortex},\ }\href@noop {} {\bibfield  {journal} {\bibinfo
  {journal} {Nature Physics}\ }\textbf {\bibinfo {volume} {15}},\ \bibinfo
  {pages} {839} (\bibinfo {year} {2019})}\BibitemShut {NoStop}%
\bibitem [{\citenamefont {Cl{\'e}ment}\ \emph {et~al.}(2017)\citenamefont
  {Cl{\'e}ment}, \citenamefont {Dehapiot}, \citenamefont {Collinet},
  \citenamefont {Lecuit},\ and\ \citenamefont {Lenne}}]{clement2017}%
  \BibitemOpen
  \bibfield  {author} {\bibinfo {author} {\bibfnamefont {R.}~\bibnamefont
  {Cl{\'e}ment}}, \bibinfo {author} {\bibfnamefont {B.}~\bibnamefont
  {Dehapiot}}, \bibinfo {author} {\bibfnamefont {C.}~\bibnamefont {Collinet}},
  \bibinfo {author} {\bibfnamefont {T.}~\bibnamefont {Lecuit}},\ and\ \bibinfo
  {author} {\bibfnamefont {P.-F.}\ \bibnamefont {Lenne}},\ }\bibfield  {title}
  {\bibinfo {title} {Viscoelastic dissipation stabilizes cell shape changes
  during tissue morphogenesis},\ }\href@noop {} {\bibfield  {journal} {\bibinfo
   {journal} {Current Biology}\ }\textbf {\bibinfo {volume} {27}},\ \bibinfo
  {pages} {3132} (\bibinfo {year} {2017})}\BibitemShut {NoStop}%
\bibitem [{\citenamefont {Cavanaugh}\ \emph
  {et~al.}(2020{\natexlab{b}})\citenamefont {Cavanaugh}, \citenamefont
  {Staddon}, \citenamefont {Banerjee},\ and\ \citenamefont
  {Gardel}}]{cavanaugh2020}%
  \BibitemOpen
  \bibfield  {author} {\bibinfo {author} {\bibfnamefont {K.~E.}\ \bibnamefont
  {Cavanaugh}}, \bibinfo {author} {\bibfnamefont {M.~F.}\ \bibnamefont
  {Staddon}}, \bibinfo {author} {\bibfnamefont {S.}~\bibnamefont {Banerjee}},\
  and\ \bibinfo {author} {\bibfnamefont {M.~L.}\ \bibnamefont {Gardel}},\
  }\bibfield  {title} {\bibinfo {title} {Adaptive viscoelasticity of epithelial
  cell junctions: From models to methods},\ }\href@noop {} {\bibfield
  {journal} {\bibinfo  {journal} {Current Opinion in Genetics \& Development}\
  }\textbf {\bibinfo {volume} {63}},\ \bibinfo {pages} {86} (\bibinfo {year}
  {2020}{\natexlab{b}})}\BibitemShut {NoStop}%
\bibitem [{\citenamefont {Odell}\ \emph {et~al.}(1981)\citenamefont {Odell},
  \citenamefont {Oster}, \citenamefont {Alberch},\ and\ \citenamefont
  {Burnside}}]{odell1981}%
  \BibitemOpen
  \bibfield  {author} {\bibinfo {author} {\bibfnamefont {G.~M.}\ \bibnamefont
  {Odell}}, \bibinfo {author} {\bibfnamefont {G.}~\bibnamefont {Oster}},
  \bibinfo {author} {\bibfnamefont {P.}~\bibnamefont {Alberch}},\ and\ \bibinfo
  {author} {\bibfnamefont {B.}~\bibnamefont {Burnside}},\ }\bibfield  {title}
  {\bibinfo {title} {The mechanical basis of morphogenesis: I. epithelial
  folding and invagination},\ }\href@noop {} {\bibfield  {journal} {\bibinfo
  {journal} {Developmental Biology}\ }\textbf {\bibinfo {volume} {85}},\
  \bibinfo {pages} {446} (\bibinfo {year} {1981})}\BibitemShut {NoStop}%
\bibitem [{\citenamefont {Munoz}\ and\ \citenamefont {Albo}(2013)}]{munoz2013}%
  \BibitemOpen
  \bibfield  {author} {\bibinfo {author} {\bibfnamefont {J.~J.}\ \bibnamefont
  {Munoz}}\ and\ \bibinfo {author} {\bibfnamefont {S.}~\bibnamefont {Albo}},\
  }\bibfield  {title} {\bibinfo {title} {Physiology-based model of cell
  viscoelasticity},\ }\href@noop {} {\bibfield  {journal} {\bibinfo  {journal}
  {Physical Review E}\ }\textbf {\bibinfo {volume} {88}},\ \bibinfo {pages}
  {012708} (\bibinfo {year} {2013})}\BibitemShut {NoStop}%
\bibitem [{\citenamefont {McFadden}\ \emph {et~al.}(2017)\citenamefont
  {McFadden}, \citenamefont {McCall}, \citenamefont {Gardel},\ and\
  \citenamefont {Munro}}]{mcfadden2017}%
  \BibitemOpen
  \bibfield  {author} {\bibinfo {author} {\bibfnamefont {W.~M.}\ \bibnamefont
  {McFadden}}, \bibinfo {author} {\bibfnamefont {P.~M.}\ \bibnamefont
  {McCall}}, \bibinfo {author} {\bibfnamefont {M.~L.}\ \bibnamefont {Gardel}},\
  and\ \bibinfo {author} {\bibfnamefont {E.~M.}\ \bibnamefont {Munro}},\
  }\bibfield  {title} {\bibinfo {title} {Filament turnover tunes both force
  generation and dissipation to control long-range flows in a model actomyosin
  cortex},\ }\href@noop {} {\bibfield  {journal} {\bibinfo  {journal} {PLoS
  Computational Biology}\ }\textbf {\bibinfo {volume} {13}},\ \bibinfo {pages}
  {e1005811} (\bibinfo {year} {2017})}\BibitemShut {NoStop}%
\bibitem [{\citenamefont {Sknepnek}\ \emph {et~al.}(2023)\citenamefont
  {Sknepnek}, \citenamefont {Djafer-Cherif}, \citenamefont {Chuai},
  \citenamefont {Weijer},\ and\ \citenamefont
  {Henkes}}]{sknepnek2021generating}%
  \BibitemOpen
  \bibfield  {author} {\bibinfo {author} {\bibfnamefont {R.}~\bibnamefont
  {Sknepnek}}, \bibinfo {author} {\bibfnamefont {I.}~\bibnamefont
  {Djafer-Cherif}}, \bibinfo {author} {\bibfnamefont {M.}~\bibnamefont
  {Chuai}}, \bibinfo {author} {\bibfnamefont {C.}~\bibnamefont {Weijer}},\ and\
  \bibinfo {author} {\bibfnamefont {S.}~\bibnamefont {Henkes}},\ }\bibfield
  {title} {\bibinfo {title} {Generating active t1 transitions through
  mechanochemical feedback},\ }\href@noop {} {\bibfield  {journal} {\bibinfo
  {journal} {eLife}\ }\textbf {\bibinfo {volume} {12}},\ \bibinfo {pages}
  {e79862} (\bibinfo {year} {2023})}\BibitemShut {NoStop}%
\bibitem [{\citenamefont {Duclut}\ \emph {et~al.}(2022)\citenamefont {Duclut},
  \citenamefont {Paijmans}, \citenamefont {Inamdar}, \citenamefont {Modes},\
  and\ \citenamefont {J{\"u}licher}}]{duclut2022active}%
  \BibitemOpen
  \bibfield  {author} {\bibinfo {author} {\bibfnamefont {C.}~\bibnamefont
  {Duclut}}, \bibinfo {author} {\bibfnamefont {J.}~\bibnamefont {Paijmans}},
  \bibinfo {author} {\bibfnamefont {M.~M.}\ \bibnamefont {Inamdar}}, \bibinfo
  {author} {\bibfnamefont {C.~D.}\ \bibnamefont {Modes}},\ and\ \bibinfo
  {author} {\bibfnamefont {F.}~\bibnamefont {J{\"u}licher}},\ }\bibfield
  {title} {\bibinfo {title} {Active t1 transitions in cellular networks},\
  }\href@noop {} {\bibfield  {journal} {\bibinfo  {journal} {The European
  Physical Journal E}\ }\textbf {\bibinfo {volume} {45}},\ \bibinfo {pages} {1}
  (\bibinfo {year} {2022})}\BibitemShut {NoStop}%
\bibitem [{\citenamefont {Bardet}\ \emph {et~al.}(2013)\citenamefont {Bardet},
  \citenamefont {Guirao}, \citenamefont {Paoletti}, \citenamefont {Serman},
  \citenamefont {L{\'e}opold}, \citenamefont {Bosveld}, \citenamefont {Goya},
  \citenamefont {Mirouse}, \citenamefont {Graner},\ and\ \citenamefont
  {Bella{\"\i}che}}]{bardet2013pten}%
  \BibitemOpen
  \bibfield  {author} {\bibinfo {author} {\bibfnamefont {P.-L.}\ \bibnamefont
  {Bardet}}, \bibinfo {author} {\bibfnamefont {B.}~\bibnamefont {Guirao}},
  \bibinfo {author} {\bibfnamefont {C.}~\bibnamefont {Paoletti}}, \bibinfo
  {author} {\bibfnamefont {F.}~\bibnamefont {Serman}}, \bibinfo {author}
  {\bibfnamefont {V.}~\bibnamefont {L{\'e}opold}}, \bibinfo {author}
  {\bibfnamefont {F.}~\bibnamefont {Bosveld}}, \bibinfo {author} {\bibfnamefont
  {Y.}~\bibnamefont {Goya}}, \bibinfo {author} {\bibfnamefont {V.}~\bibnamefont
  {Mirouse}}, \bibinfo {author} {\bibfnamefont {F.}~\bibnamefont {Graner}},\
  and\ \bibinfo {author} {\bibfnamefont {Y.}~\bibnamefont {Bella{\"\i}che}},\
  }\bibfield  {title} {\bibinfo {title} {Pten controls junction lengthening and
  stability during cell rearrangement in epithelial tissue},\ }\href@noop {}
  {\bibfield  {journal} {\bibinfo  {journal} {Developmental cell}\ }\textbf
  {\bibinfo {volume} {25}},\ \bibinfo {pages} {534} (\bibinfo {year}
  {2013})}\BibitemShut {NoStop}%
\bibitem [{\citenamefont {Kim}\ \emph {et~al.}(2021)\citenamefont {Kim},
  \citenamefont {Pochitaloff}, \citenamefont {Stooke-Vaughan},\ and\
  \citenamefont {Camp{\`a}s}}]{kim2021}%
  \BibitemOpen
  \bibfield  {author} {\bibinfo {author} {\bibfnamefont {S.}~\bibnamefont
  {Kim}}, \bibinfo {author} {\bibfnamefont {M.}~\bibnamefont {Pochitaloff}},
  \bibinfo {author} {\bibfnamefont {G.~A.}\ \bibnamefont {Stooke-Vaughan}},\
  and\ \bibinfo {author} {\bibfnamefont {O.}~\bibnamefont {Camp{\`a}s}},\
  }\bibfield  {title} {\bibinfo {title} {Embryonic tissues as active foams},\
  }\href@noop {} {\bibfield  {journal} {\bibinfo  {journal} {Nature Physics}\
  }\textbf {\bibinfo {volume} {17}},\ \bibinfo {pages} {859} (\bibinfo {year}
  {2021})}\BibitemShut {NoStop}%
\bibitem [{\citenamefont {Noll}\ \emph {et~al.}(2017)\citenamefont {Noll},
  \citenamefont {Mani}, \citenamefont {Heemskerk}, \citenamefont {Streichan},\
  and\ \citenamefont {Shraiman}}]{noll2017active}%
  \BibitemOpen
  \bibfield  {author} {\bibinfo {author} {\bibfnamefont {N.}~\bibnamefont
  {Noll}}, \bibinfo {author} {\bibfnamefont {M.}~\bibnamefont {Mani}}, \bibinfo
  {author} {\bibfnamefont {I.}~\bibnamefont {Heemskerk}}, \bibinfo {author}
  {\bibfnamefont {S.~J.}\ \bibnamefont {Streichan}},\ and\ \bibinfo {author}
  {\bibfnamefont {B.~I.}\ \bibnamefont {Shraiman}},\ }\bibfield  {title}
  {\bibinfo {title} {Active tension network model suggests an exotic mechanical
  state realized in epithelial tissues},\ }\href@noop {} {\bibfield  {journal}
  {\bibinfo  {journal} {Nature Physics}\ }\textbf {\bibinfo {volume} {13}},\
  \bibinfo {pages} {1221} (\bibinfo {year} {2017})}\BibitemShut {NoStop}%
\bibitem [{\citenamefont {Gustafson}\ \emph
  {et~al.}(2022{\natexlab{b}})\citenamefont {Gustafson}, \citenamefont
  {Claussen}, \citenamefont {De~Renzis},\ and\ \citenamefont
  {Streichan}}]{gustafson2022}%
  \BibitemOpen
  \bibfield  {author} {\bibinfo {author} {\bibfnamefont {H.~J.}\ \bibnamefont
  {Gustafson}}, \bibinfo {author} {\bibfnamefont {N.}~\bibnamefont {Claussen}},
  \bibinfo {author} {\bibfnamefont {S.}~\bibnamefont {De~Renzis}},\ and\
  \bibinfo {author} {\bibfnamefont {S.~J.}\ \bibnamefont {Streichan}},\
  }\bibfield  {title} {\bibinfo {title} {Patterned mechanical feedback
  establishes a global myosin gradient},\ }\href@noop {} {\bibfield  {journal}
  {\bibinfo  {journal} {Nature Communications}\ }\textbf {\bibinfo {volume}
  {13}},\ \bibinfo {pages} {7050} (\bibinfo {year}
  {2022}{\natexlab{b}})}\BibitemShut {NoStop}%
\bibitem [{\citenamefont {Bi}\ \emph {et~al.}(2015)\citenamefont {Bi},
  \citenamefont {Lopez}, \citenamefont {Schwarz},\ and\ \citenamefont
  {Manning}}]{bi2015}%
  \BibitemOpen
  \bibfield  {author} {\bibinfo {author} {\bibfnamefont {D.}~\bibnamefont
  {Bi}}, \bibinfo {author} {\bibfnamefont {J.}~\bibnamefont {Lopez}}, \bibinfo
  {author} {\bibfnamefont {J.~M.}\ \bibnamefont {Schwarz}},\ and\ \bibinfo
  {author} {\bibfnamefont {M.~L.}\ \bibnamefont {Manning}},\ }\bibfield
  {title} {\bibinfo {title} {A density-independent rigidity transition in
  biological tissues},\ }\href@noop {} {\bibfield  {journal} {\bibinfo
  {journal} {Nature Physics}\ }\textbf {\bibinfo {volume} {11}},\ \bibinfo
  {pages} {1074} (\bibinfo {year} {2015})}\BibitemShut {NoStop}%
\bibitem [{\citenamefont {Damavandi}\ \emph {et~al.}(2022)\citenamefont
  {Damavandi}, \citenamefont {Lawson-Keister},\ and\ \citenamefont
  {Manning}}]{damavandi2022universal}%
  \BibitemOpen
  \bibfield  {author} {\bibinfo {author} {\bibfnamefont {O.~K.}\ \bibnamefont
  {Damavandi}}, \bibinfo {author} {\bibfnamefont {E.}~\bibnamefont
  {Lawson-Keister}},\ and\ \bibinfo {author} {\bibfnamefont {M.~L.}\
  \bibnamefont {Manning}},\ }\bibfield  {title} {\bibinfo {title} {Universal
  features of rigidity transitions in vertex models for biological tissues},\
  }\bibfield  {journal} {\bibinfo  {journal} {bioRxiv}\ }\href
  {https://doi.org/10.1101/2022.06.01.494406} {10.1101/2022.06.01.494406}
  (\bibinfo {year} {2022})\BibitemShut {NoStop}%
\bibitem [{\citenamefont {Gibson}\ \emph {et~al.}(2006)\citenamefont {Gibson},
  \citenamefont {Patel}, \citenamefont {Nagpal},\ and\ \citenamefont
  {Perrimon}}]{gibson2006emergence}%
  \BibitemOpen
  \bibfield  {author} {\bibinfo {author} {\bibfnamefont {M.~C.}\ \bibnamefont
  {Gibson}}, \bibinfo {author} {\bibfnamefont {A.~B.}\ \bibnamefont {Patel}},
  \bibinfo {author} {\bibfnamefont {R.}~\bibnamefont {Nagpal}},\ and\ \bibinfo
  {author} {\bibfnamefont {N.}~\bibnamefont {Perrimon}},\ }\bibfield  {title}
  {\bibinfo {title} {The emergence of geometric order in proliferating metazoan
  epithelia},\ }\href@noop {} {\bibfield  {journal} {\bibinfo  {journal}
  {Nature}\ }\textbf {\bibinfo {volume} {442}},\ \bibinfo {pages} {1038}
  (\bibinfo {year} {2006})}\BibitemShut {NoStop}%
\bibitem [{\citenamefont {Banerjee}\ \emph {et~al.}(2015)\citenamefont
  {Banerjee}, \citenamefont {Utuje},\ and\ \citenamefont
  {Marchetti}}]{banerjee2015}%
  \BibitemOpen
  \bibfield  {author} {\bibinfo {author} {\bibfnamefont {S.}~\bibnamefont
  {Banerjee}}, \bibinfo {author} {\bibfnamefont {K.~J.}\ \bibnamefont
  {Utuje}},\ and\ \bibinfo {author} {\bibfnamefont {M.~C.}\ \bibnamefont
  {Marchetti}},\ }\bibfield  {title} {\bibinfo {title} {Propagating stress
  waves during epithelial expansion},\ }\href@noop {} {\bibfield  {journal}
  {\bibinfo  {journal} {Physical Review Letters}\ }\textbf {\bibinfo {volume}
  {114}},\ \bibinfo {pages} {228101} (\bibinfo {year} {2015})}\BibitemShut
  {NoStop}%
\bibitem [{\citenamefont {Hino}\ \emph {et~al.}(2020)\citenamefont {Hino},
  \citenamefont {Rossetti}, \citenamefont {Mar{\'\i}n-Llaurad{\'o}},
  \citenamefont {Aoki}, \citenamefont {Trepat}, \citenamefont {Matsuda},\ and\
  \citenamefont {Hirashima}}]{hino2020}%
  \BibitemOpen
  \bibfield  {author} {\bibinfo {author} {\bibfnamefont {N.}~\bibnamefont
  {Hino}}, \bibinfo {author} {\bibfnamefont {L.}~\bibnamefont {Rossetti}},
  \bibinfo {author} {\bibfnamefont {A.}~\bibnamefont
  {Mar{\'\i}n-Llaurad{\'o}}}, \bibinfo {author} {\bibfnamefont
  {K.}~\bibnamefont {Aoki}}, \bibinfo {author} {\bibfnamefont {X.}~\bibnamefont
  {Trepat}}, \bibinfo {author} {\bibfnamefont {M.}~\bibnamefont {Matsuda}},\
  and\ \bibinfo {author} {\bibfnamefont {T.}~\bibnamefont {Hirashima}},\
  }\bibfield  {title} {\bibinfo {title} {Erk-mediated mechanochemical waves
  direct collective cell polarization},\ }\href@noop {} {\bibfield  {journal}
  {\bibinfo  {journal} {Developmental Cell}\ }\textbf {\bibinfo {volume}
  {53}},\ \bibinfo {pages} {646} (\bibinfo {year} {2020})}\BibitemShut
  {NoStop}%
\bibitem [{\citenamefont {Wang}\ \emph {et~al.}(2020)\citenamefont {Wang},
  \citenamefont {Merkel}, \citenamefont {Sutter}, \citenamefont
  {Erdemci-Tandogan}, \citenamefont {Manning},\ and\ \citenamefont
  {Kasza}}]{wang2020anisotropy}%
  \BibitemOpen
  \bibfield  {author} {\bibinfo {author} {\bibfnamefont {X.}~\bibnamefont
  {Wang}}, \bibinfo {author} {\bibfnamefont {M.}~\bibnamefont {Merkel}},
  \bibinfo {author} {\bibfnamefont {L.~B.}\ \bibnamefont {Sutter}}, \bibinfo
  {author} {\bibfnamefont {G.}~\bibnamefont {Erdemci-Tandogan}}, \bibinfo
  {author} {\bibfnamefont {M.~L.}\ \bibnamefont {Manning}},\ and\ \bibinfo
  {author} {\bibfnamefont {K.~E.}\ \bibnamefont {Kasza}},\ }\bibfield  {title}
  {\bibinfo {title} {Anisotropy links cell shapes to tissue flow during
  convergent extension},\ }\href@noop {} {\bibfield  {journal} {\bibinfo
  {journal} {Proceedings of the National Academy of Sciences}\ }\textbf
  {\bibinfo {volume} {117}},\ \bibinfo {pages} {13541} (\bibinfo {year}
  {2020})}\BibitemShut {NoStop}%
\end{thebibliography}
\end{document}


\setstcolor{red}

\title{{\bf Supplemental Material}\\ Tension remodeling controls topological transitions in epithelial tissues}
\author[1]{Fernanda P\'erez-Verdugo}
\author[1]{Shiladitya Banerjee}
\affil[1]{ Department of Physics, Carnegie Mellon University, Pittsburgh, PA 15213, USA}

\date{}
\maketitle

\section{Simulation Methods}

\subsection{Initial configuration}
\begin{figure}[htp!] 
\centering 
 \includegraphics[width=.7\linewidth]{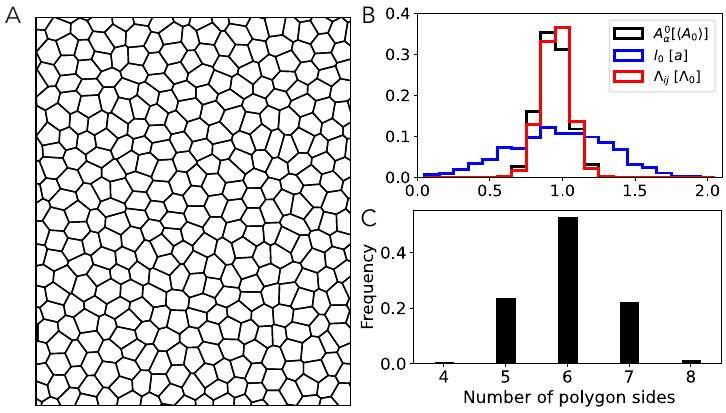} 
\caption{\textbf{Initial configuration of the tissue}. (A) Representative section of the energy relaxed tissue. (B) Histograms of cell target area, junction rest length, and junction tension at the relaxed initial state. Rest length is expressed in units of $\textsl{a}$, which corresponds to the side of a regular hexagon with area $\langle A_0\rangle$ $\left(\textsl{a} = 0.62\sqrt{\langle A_0\rangle}\right)$. (C) Histogram of polygon sidedness in the initial configuration.}
\label{fig.FigS1}
\end{figure}

We created a disordered tissue composed of 494 cells in a box of dimensions $L_x \times L_y$ (Table~\ref{tab.parameters}), with periodic boundary conditions. The disordered tissue was built via Voronoi
tessellation, where the positions of the cell centers were generated
by a Monte Carlo simulation of hard disks, with an area fraction equal to 0.71 \cite{perez2020vertex}. On this disordered tissue comprising variable-sided polygons, we obtained the energy relaxed state by evolving the system using the standard vertex model Hamiltonian:
\begin{equation}
    E=\frac{1}{2}K \sum_\alpha \left(A_\alpha - A_\alpha^0\right)^2 + \frac{1}{2} K_P \sum_\alpha \left(P_\alpha - P_\alpha^0\right)^2,
\end{equation}
where $P_\alpha$ and $P_\alpha^0=P_0$ define the cellular perimeters and their target values, respectively. We used $K=1, K_P=0.2, P_0=3.5$, and $A_\alpha^0$ was drawn from a normal distribution with unit mean and standard deviation 0.1. During the initial energy relaxation process, we allowed T1 rearrangements when cellular edges became smaller than a length threshold $l_{T_1}=0.05$ and if the topology change decreased the overall tissue energy. This procedure ensured that the previous relaxed tissue configuration is in a solid state \cite{bi2015} containing tricellular vertices only.
Using the configuration of the resulting tissue, we initiated a second round of energy relaxation using the hamiltonian defined in Eq. (1) of the main text, with active terms set to zero. During this process, we preserved the values of the target areas, and assumed that the initial rest length $l_{ij}^0$ is equal to the initial junction length $l_{ij}$. Further, cell junctions were assigned tension values $\Lambda_{ij}$, drawn from a uniform distribution with mean 0.1 and standard deviation 0.01. The parameter values are given in Table~\ref{tab.parameters}. We let this tissue relax its energy by setting $k_E=k_C=0$, $\sigma=0$, $\Gamma_a=0$. As a result, the tensions $\Lambda_{ij}$ did not change, while $l_{ij}^0$ and $l_{ij}$ reach approximately normal distributions with mean $0.62$ (side of a regular hexagon with unit area), and standard deviation $\sim 0.2$. The relaxed state, defining the initial configuration of our simulations, represents a solid-like tissue \cite{farhadifar2007influence} with tricellular vertices only, and polygonal shapes from squares to octagons (Fig.~\ref{fig.FigS1}). Note that this relaxed state of the tissue could have been obtained only using the second relaxation step. However, our two-step method minimizes the noise in the system arising due to the creation and resolution of 4-fold vertices, making the relaxation process much faster.

\newpage 
\begin{table}
\begin{center}
\begin{tabular}{ |c|c|c| } 
\hline
Parameter & Symbol & Value  \\
\hline
Area elastic modulus & $K$ & 1  \\ 
Mean preferred area & $\langle A_{\alpha}^0\rangle $ & 1  \\ 
Friction coefficient & $\mu$ & 0.2 ($\SI{28}{\second}$) \\ 
Mean initial tension & $\Lambda_0$& 0.1   \\ 
Tension in newly created junction & $\Lambda_\text{birth}$& 0.1   \\ 
Length of a newly created junction & $l_\text{birth}$ & 1.5 $l_{T_1}$  \\ 
Active contractility & $\Gamma_a$& 0.03   \\ 
Noise amplitude & $\sigma$ & 0.02 \\
Simulation box length & $L_x$ & $\sim 20$ \\ 
Simulation box width & $L_y$ & $\sim 24$  \\ 
Strain relaxation rate & $k_L$& 1   \\ 
Tension remodeling rate under contraction & $k_C$& $\in \left[0.02,0.23\right]$   \\ 
Tension remodeling rate under extension & $k_E$& $\in \left[0.02,0.23\right]$   \\
Critical strain threshold & $\epsilon_c$& 0.1   \\ 
Length threshold for attempting T1 transitions & $l_{T_1}$& 0.05   \\ 
Time between attempts to resolve 4-fold vertices & $\tau_{\text{test}}$ & 0.04   \\ 
Tension relaxation timescale & $\tau_{\Lambda}$ & 10  \\
Relaxation timescale for tension fluctuations & $\tau$ & 0.4  \\
Integration time step & $\Delta t$ & 0.004  \\
\hline
\end{tabular}
\end{center}
\caption{\label{tab.parameters}Model parameters.}
\end{table}

\subsection{Rules for T1 transitions and higher-order vertex assembly}

\begin{figure}[h!] 
\centering 
\includegraphics[width=\linewidth]{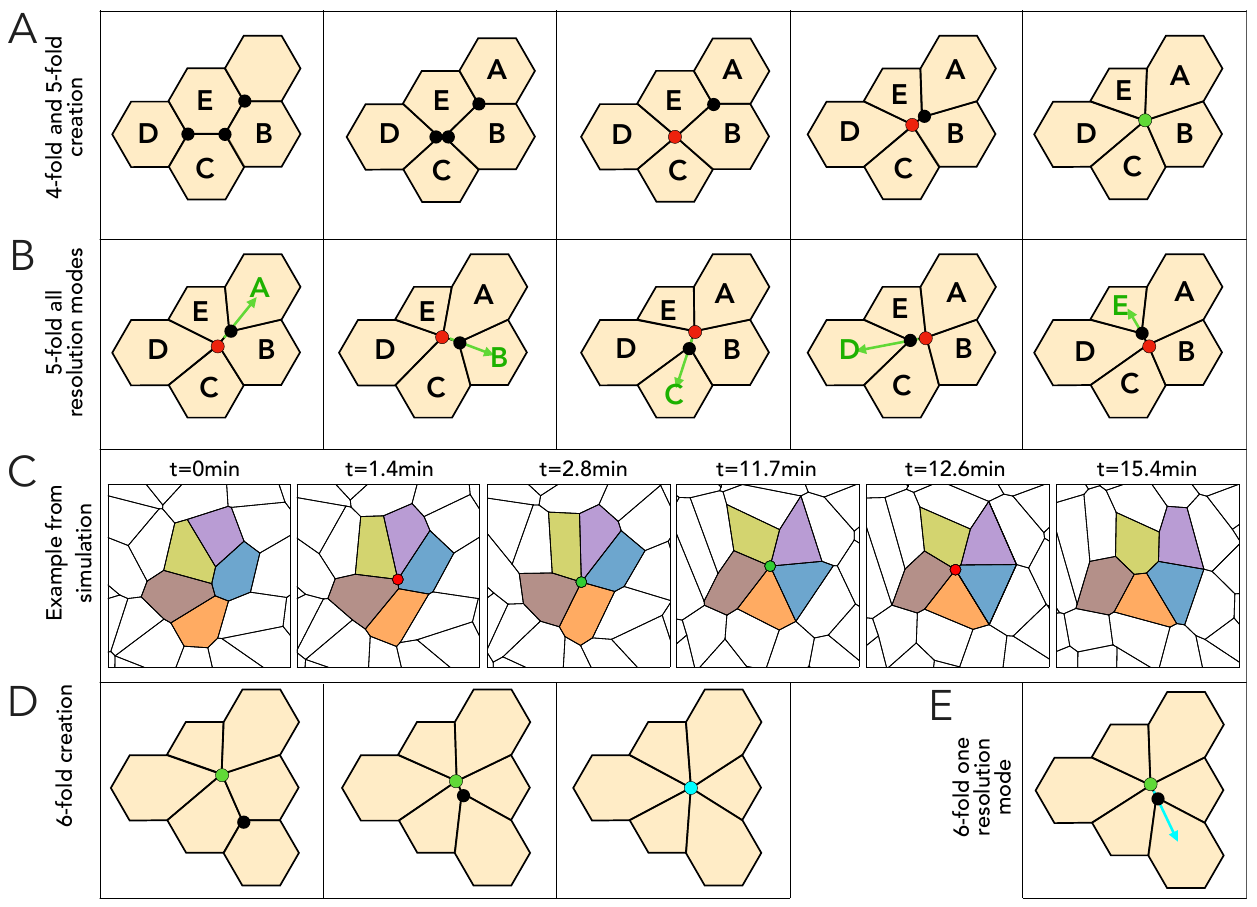} 
\caption{\textbf{Higher-order vertices creation and resolution.} (A) 4-fold and 5-fold creation. (B) Five modes of resolution of a 5-fold vertex. (C) Example of the creation and resolution of a 5-fold vertex (from Movie 4). (D) 6-fold creation from the merging of a 5-fold and a 3-fold vertices. (E) One (out of six) modes of resolution of a 6-fold vertex. Colores circles represent 3-fold (black), 4-fold (red), 5-fold (green), and 6-fold (cyan) vertices.} 
\label{fig.FigS2}
\end{figure}

\begin{enumerate}
\item If a junction shared by a $n$-fold ($n\geq3$) vertex $i$ and a 3-fold vertex $j$, with total tension $\Lambda_{ij}$, shrinks below a threshold length $l_{ij} < l_{T_1}$, then we remove one of the vertices ($j$), while transforming the other ($i$) into a $\left( n+1\right)$-fold vertex. See Fig.~\ref{fig.FigS2}A-C for examples for $n=\lbrace 3,4,5\rbrace.$ During this process, each shoulder junction sustaining the $\left( n+1\right)$-fold vertex gains $1/\left( n+1\right)$ of the tension ($\Lambda_{ij}$) in the deleted junction, which remains in the system until the $\left( n+1\right)$-fold is resolved. Movie 4 shows a system is which 3-fold, 4-fold and 5-fold vertices are allowed.\\

Note: In the main text simulations we only show 3-fold and 4-fold vertices. In those simulations, if a junction shared by two vertices $i$ and $j$ shrinks below a threshold $l_{ij} < l_{T_1}$, and at least one of them is a 4-fold vertex, no higher-order vertex is created, and the tension remodeling rate $k_C$ is set to zero until $l_{ij}$ reaches the threshold $l_{T_1}$.

 \item Every $\tau_{\text{test}}$ timesteps, we attempt to resolve the $\left( n+1\right)$-fold vertices present in the tissue. The resolution is tested for each of the $\left( n+1\right)$ configurations. During the $n$-fold vertex resolution timestep, we create a junction of length $1.5l_{T_1}$ in the direction $\left(\VEC R_c^{\alpha} - \VEC r_{n+1}\right)/\mid \VEC R_c^{\alpha} - \VEC r_{n+1}\mid$, where $\VEC R_c^{\alpha}$ is the center of the cell $\alpha$ surrounding the $(n+1)$-fold vertex with position $\VEC r_{n+1}$. See Fig.~\ref{fig.FigS2}B-D for examples with $n=\lbrace4,5\rbrace.$ We assign a tension value $\Lambda_{ij}$ for the newly created junction drawn from a normal distribution with mean value $\Lambda_0$ and standard deviation $0.1\Lambda_0$. Then, the total tension in the new junction is given by $\Lambda_{\text{birth}} = \Lambda_{ij}+1.5\Gamma_a l_{T_1}$. The rest length $l_{ij}^0$ of the new junction is drawn from a truncated (only positive values) normal distribution with mean value $l_{T_1}$ and standard deviation equals $0.1l_{T_1}$. During this process, each shoulder junction sustaining the newly created junction loses $1/\left( n+1\right)$ of $\Lambda_{\text{birth}}$. Finally, we calculate the forces at vertices $i$ and $j$ for each configuration.\\
 
 Note: For the 4-fold vertex resolution we combine the four possible configurations into two - the original (reverse T1 transition) and the perpendicular (T1 transition) one. The original one is defined by the direction $\left(\VEC R_c^{B} - \VEC R_c^{D}\right)/\mid \VEC R_c^{B} - \VEC R_c^{D}\mid$, and the perpendicular by $\left(\VEC R_c^{E} - \VEC R_c^{C}\right)/\mid \VEC R_c^{E} - \VEC R_c^{C}\mid$, with $\lbrace S,E,B,C\rbrace$ as shown in Fig.~\ref{fig.FigS2}A.

 \item  If the effective force between $i$ and $j$ is attractive in all the tested configurations, then the $\left( n+1\right)$-fold vertex is considered stable, and we proceed to delete $j$ again. However, if the effective force between $i$ and $j$ is repulsive, at least in one case, then the $\left( n+1\right)$-fold vertex is unstable. Therefore, we choose the configuration with the largest repulsive force between the $n$-fold $i$ and the 3-fold $j$ vertices.
 
\end{enumerate}

\subsection{Choice of model parameters}
Our model involves approximately 19 parameters, as listed in Table 1. Many of these parameters are calibrated from prior experimental studies and measurements, several parameters have been varied in our simulations, and the remainder can be eliminated through non-dimensionalization of the equations of motion. Below, we provide a detailed account of our parameter choices.

Within the set of listed parameters, two pertain to defining tissue size ($L_x$ and $L_y$), one sets the integration time step ($\Delta t$), and another establishes the frequency of attempting T1 transitions ($\tau_\text{test}$). These parameters are essentially model-specific choices, akin to those in any numerical work. Out of the remaining fifteen parameters, three of them (area elastic modulus $K$, mean preferred area $\langle A_\alpha^0\rangle$, and friction coefficient $\mu$) have been used to non-dimensionalize the equations of motion. Specifically, we non-dimensionalized force scales by $K(A_\alpha^0)^{3/2}$, length scales by $\sqrt{A_\alpha^0}$, time scales by $\mu/KA_\alpha^0$, setting $K=1$, $\langle A_\alpha^0\rangle=1$, and $\mu=0.2$ ($\sim\SI{28}{\second}$), where $\langle ..\rangle$ represents population average.  The positive value for the mean initial tension, $\langle \Lambda_{ij}\rangle=\Lambda_0=0.1$, has been chosen to ensure that the initial state of the non-active tissue (further details are available in Section 1.1) resembles a stable, solid-like network~\cite{farhadifar2007influence}. Similar values have been adopted in previous studies that utilized the tension-remodeling model at the junction level to fit experimental data~\cite{staddon2019mechanosensitive,cavanaugh2020rhoa,nishizawa2023two}.

The tension components arising from contractility ($\Gamma_a$) and fluctuations ($\sigma$) are assumed to be considerably smaller than the mean initial tension. The relaxation timescale for tension fluctuations is assumed to be twice the friction value, ensuring the persistence of this stochastic term during short time intervals. The longest timescale is attributed to tension relaxation, $\tau_\Lambda$, which facilitates the dominance of tension remodeling dynamics during the temporal window encompassing 4-fold vertex formation, followed by instantaneous resolution or brief stalling periods.

Out of the remaining seven parameters, four are closely associated with tension remodeling dynamics: strain relaxation rate ($k_L$), critical strain threshold ($\varepsilon_c$), and two tension remodeling rates ($k_E$ and $k_C$). For the first two parameters, we have chosen values of similar magnitude as previously benchmarked in references~\cite{staddon2019mechanosensitive,cavanaugh2020rhoa}, derived from experimental data. Regarding the tension remodeling rates, we explored values that yield stable systems, as expounded upon in Fig.~1 of the main text.

It is crucial to acknowledge that the values employed in previous studies fall within the range of our examined parameter space. Lastly, we acknowledge that direct experimental data for the tension and length of newly created junctions, as well as the length threshold for 4-fold vertex formation, are exceptionally challenging to obtain due to resolution limitations in current imaging techniques. Nonetheless, we provide informed estimations for these parameters and subsequently vary these three parameters, along with the tension resetting rules during 4-fold creation and resolution, demonstrating that our results remain robust (Figs.~S10-S12).


\section{Analysis of a positive feedback model between junction tension and strain}

\begin{figure}[h!] 
\centering 
\includegraphics[width=.6\linewidth]{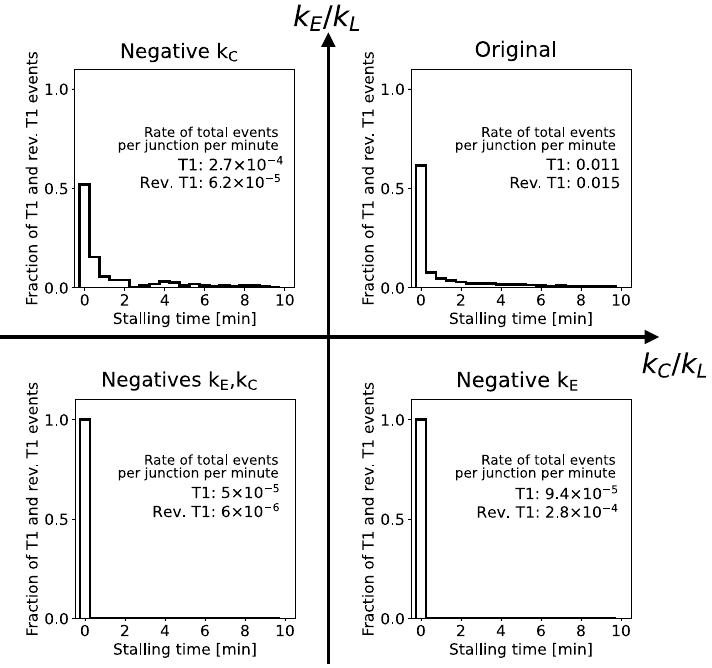} 
\caption{\textbf{Transiently stable 4-fold vertices rely on tension increasing during contraction}. Histograms of the stalling time for T1 and reverse T1 events, for: 
 original simulation $\left(k_C/k_L=0.1, k_E/k_L=0.2\right)$; negative $k_C$ $\left(k_C/k_L=-0.1, k_E/k_L=0.2\right)$; negatives $k_E,k_C$ $\left(k_C/k_L=-0.1, k_E/k_L=-0.2\right)$; and, negative $k_E$ $\left(k_C/k_L=0.1, k_E/k_L=-0.2\right)$.}
\label{fig.FigS3}
\end{figure}

In the main text, we studied a negative feedback model between junction tension and strain. Specifically, we used the model that tension increases in junctions under contraction, with a rate proportional to $k_C\ (\geq0)$, while it decreases in elongated junctions at a rate proportional to $k_E\ (\geq0)$. Here, we analyze the cases where the rates $k_C$ and $k_E$ can take negative values, signifying a positive feedback between tension and strain. We find that transiently stable 4-fold vertices can arise for $k_C<0$, if $k_E$ is positive and large enoughn (Fig.~\ref{fig.FigS3}). A positive large value for $k_E$ is required to decrease the tension in the extending shoulder junctions compared to the tension in the contracting junction during a T1 event, in order to stabilize a 4-fold vertex. However, with $k_C<0$, the frequency of T1 events is much lower. For negative values of $k_E$, all T1 transitions occur instantaneously and 4-fold vertices are unstable. 

\section{Mean-field model}
\begin{figure}[b!] 
\centering 
 \includegraphics[width=\linewidth]{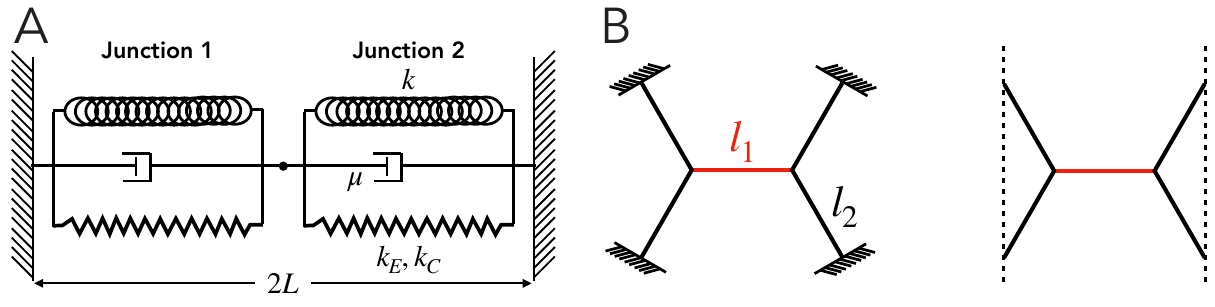} 
\caption{\textbf{Schematic of the mean-field models}. (A) Effective system composed of two junctions of natural length $2L$, under fixed boundary conditions. Each junction is composed of an elastic element with spring constant $k$, a dashpot with friction coefficient $\mu$, and a tension $\Lambda$ that remodels at a rate $k_C$ under contraction, and $k_E$ under stretch. Junction rest length remodels at a rate $k_L$. (B) Schematic of an effective five-junction-system in two dimensions, as part of a hexagonal lattice. The central junction is activated by contraction, sustained by four shoulder junctions. (Left) Shoulder junctions are under a fixed boundary condition. (Right) The four shoulder junctions are free to move vertically.}
\label{fig.FigS4}
\end{figure}

To analytically predict the mechanical stability of cell junctions under contraction, we consider an effective medium theory of the system consisting of two cell junctions in series, connected in parallel to an effective elastic medium of spring constant $k$ (Fig.~\ref{fig.FigS4}A). Tension in each junction $i$ ($i=1,2$) is given by $\Lambda_i$, with length $l_i$, rest length $l_{0i}$, deforming against an overdamped medium with friction coefficient $\mu$. We choose to activate junction 1 with contractility $\Gamma_a>0$, neglect tension fluctuations, and set the critical strain threshold to 0 for simplicity. Junction tensions and rest lengths evolve following the dynamics defined in main text,
\begin{align}
\frac{{\rm d}\Lambda_{1}}{{\rm d}t}&=-k_C(l_1-l_{01}),\label{eq.model_junction2}\\
\frac{{\rm d}l_{01}}{{\rm d}t}&= -k_L(l_{01}-l_1),\label{eq.model_junction3}\\
\frac{{\rm d}\Lambda_{2}}{{\rm d}t}&=-k_E(l_2-l_{02}),\label{eq.model_junction4}\\
\frac{{\rm d}l_{02}}{{\rm d}t}&= -k_L(l_{02}-l_2).
\label{eq.model_junction5}
\end{align}

Assuming the system conserve its total length, we have the constraint $2L=l_1+l_2$. Dynamics of junction length then follows from considering the force-balance equation at the vertex between the two junctions,
\begin{align}
2\mu \frac{{\rm d} l_1}{{\rm d}t}&= k(l_2-L)+k(L-l_1)+\Lambda_2-\Lambda_1-\Gamma_a l_1,\nonumber\\
&= k(2L-l_1-L)+k(L-l_1)+\Lambda_2-\Lambda_1-\Gamma_a l_1,\nonumber\\
&= 2k(L-l_1)+\Lambda_2-\Lambda_1-\Gamma_a l_1.
\label{eq.lpunto1}
\end{align}

\subsection{Mechanical stability of cell junctions}
The activated junction (junction 1) will collapse to zero length if the system is unstable to contraction. Stability is thus defined by the junction reaching a non-zero length $l_1>0$ at steady-state. At steady-state, solution to Eqs.~\eqref{eq.model_junction2}-\eqref{eq.lpunto1} are given by,
\begin{align}
0&=2kL-(2k+\Gamma_a )l_1^{\text{EQ}}+\Lambda_2^{\text{EQ}}-\Lambda_1^{\text{EQ}},\label{eq.l1EQ} \\
l_{01}^{\text{EQ}}&=l_{1}^{\text{EQ}},\label{eq.l01EQ}\\
l_{02}^{\text{EQ}}&=l_{2}^{\text{EQ}}= 2L-l_{1}^{\text{EQ}}.
\label{eq.l02EQ}
\end{align}
Since ${\rm d} \Lambda_1/{\rm d}t = -(k_C/k_L){\rm d} l_{01}/{\rm d}t$, and ${\rm d} \Lambda_2/{\rm d}t = -(k_C/k_L){\rm d} l_{02}/{\rm d}t$, we can integrate these equations from $t=0$ to the time at which steady-state is reached, to calculate the steady-state tension values $\Lambda_1^{\text{EQ}}$ and $\Lambda_2^{\text{EQ}}$. Considering that initial lengths and tensions were given by $L$, and $T_0$, respectively, we obtain
\begin{align}
\Lambda_1^{\text{EQ}} &= T_0 - \frac{k_C}{k_L}\left(l_{01}^{\text{EQ}} - L\right),\label{eq.TEQ1}\\
\Lambda_2^{\text{EQ}} &= T_0 - \frac{k_E}{k_L}\left(l_{02}^{\text{EQ}} - L\right).
\label{eq.TEQ2}
\end{align}
Using relations in Eqs.~\eqref{eq.l01EQ}-\eqref{eq.TEQ2} in Eq.~\eqref{eq.l1EQ} we get,
\begin{align}
0&=2kL-(2k+\Gamma_a )l_1^{\text{EQ}}+T_0 - \frac{k_E}{k_L}\left(l_{02}^{\text{EQ}} - L\right)-T_0 + \frac{k_C}{k_L}\left(l_{01}^{\text{EQ}} - L\right), \nonumber \\
0&=2kL-(2k+\Gamma_a )l_1^{\text{EQ}}- \frac{k_E}{k_L}\left(2L-l_{1}^{\text{EQ}} - L\right) + \frac{k_C}{k_L}\left(l_{1}^{\text{EQ}} - L\right),\nonumber\\
0&=2kL-(2k+\Gamma_a )l_1^{\text{EQ}}- \frac{k_E}{k_L}\left(L-l_{1}^{\text{EQ}}\right) - \frac{k_C}{k_L}\left(L-l_{1}^{\text{EQ}} \right),\nonumber\\
0&=2kL-(2k+\Gamma_a )l_1^{\text{EQ}}- \left(\frac{k_E+k_C}{k_L}\right)L+ \left(\frac{k_E+k_C}{k_L}\right)l_{1}^{\text{EQ}},\nonumber\\
0&=L\left(2k - \frac{k_E+k_C}{k_L} \right)-\left(2k+\Gamma_a -\frac{k_E+k_C}{k_L}\right)l_1^{\text{EQ}}.
\label{eq.l1eQre}
\end{align}
The above equation gives,
\begin{align}
l_1^{\text{EQ}} &=L\frac{\left(2k - \frac{k_E+k_C}{k_L} \right)}{\left(2k+\Gamma_a -\frac{k_E+k_C}{k_L}\right)}, 
\label{eq.l1eQf}
\end{align}
Thus, in order to have $L>l_1^{\text{EQ}}>0$,  $(k_E+k_C)/k_L$ has to be smaller than $2k$. 

\subsection{Tension change due to remodeling}
In the main text, we presented results relating the stability of four-fold vertices to reduction in tension in the tissue. Here we derive the condition for reduction in tension using the effective medium model. We can write Eqs.~\eqref{eq.TEQ1} and \eqref{eq.TEQ2} in a more general form, with the initial lengths of a junction under contraction (stretch) given by $L_C^{\text{ini}}$ ($L_E^{\text{ini}}$),
\begin{align}
\Lambda_C^{\text{EQ}} &= T_0 - \frac{k_C}{k_L}\left(l_{0C}^{\text{EQ}} - L_C^{\text{ini}}\right),\label{eq.TEQ1GEN}\\
\Lambda_E^{\text{EQ}} &= T_0 - \frac{k_E}{k_L}\left(l_{0E}^{\text{EQ}} - L_E^{\text{ini}}\right).
\label{eq.TEQ2GEN}
\end{align}
Assuming that we have $A$ junctions under contraction and $B$ junction under stretch, all of them with initial tension $T_0$, the global change in tissue tension from an undeformed state is given by 
\begin{align}
\Delta \Lambda &= \left[ A T_0 - \frac{k_C}{k_L} \sum_{C=1}^A \left(l_{0C}^{\text{EQ}} - L_C^{\text{ini}}\right) + B T_0 - \frac{k_E}{k_L} \sum_{E=1}^B \left(l_{0E}^{\text{EQ}} - L_E^{\text{ini}}\right)\right] - (A+B)T_0,\nonumber\\
&= - \frac{k_C}{k_L} \sum_{C=1}^A \left(l_{0C}^{\text{EQ}} - L_C^{\text{ini}}\right)- \frac{k_E}{k_L} \sum_{E=1}^B \left(l_{0E}^{\text{EQ}} - L_E^{\text{ini}}\right),\nonumber\\
&=  -\left[\frac{k_E}{k_L} \sum_{E=1}^B \left(l_{0E}^{\text{EQ}} - L_E^{\text{ini}}\right)-\frac{k_C}{k_L} \sum_{C=1}^A \left( L_C^{\text{ini}}-l_{0C}^{\text{EQ}}\right)\right].
\label{eq.dTglobal}
\end{align}
We then define $\delta L^+ = \sum_{E=1}^B \left(l_{0E}^{\text{EQ}} - L_E^{\text{ini}}\right)$, and  $\delta L^- = \sum_{C=1}^A \left(L_C^{\text{ini}}-l_{0C}^{\text{EQ}}\right)$, where $\delta L^+$ represents the net elongation of the junctions under stretch, while $\delta L^-$ represents the net contraction of the junctions under contraction. If $\delta L^+ = \beta \delta L^-$, we then have
\begin{align}
\Delta \Lambda &=  -\left(\frac{k_E}{k_L} \delta L^+ -\frac{k_C}{k_L} \delta L^- \right),\nonumber\\
&=  -\left(\frac{k_E}{k_L}\beta \delta L^- -\frac{k_C}{k_L} \delta L^- \right),\nonumber\\
&=  -\frac{\delta L^-}{k_L}\left(\beta k_E - k_C\right),
\label{eq.dTglobaldelta}
\end{align}
From the last equation, we can see that the condition for reducing the global tension is $\beta k_E>k_C$, with $\beta$ depending on the increase or decrease of the total junction length after activation.
In a system that conserves its total junction length ($\delta L^+ = \delta L^-$), the condition for reducing the global tension is given by $k_E>k_C$. However, if after the activation, the system increases its total junction length ($\delta L^+ > \delta L^-$), then the condition for reducing tension is given by $\beta k_E>k_C$, with $\beta>1$. In the context of a confluent tissue simulated with the vertex model, a solid tissue tries to maintain the initial steady-state configuration, and we expect the condition $k_E>k_C$ to hold for global tension reduction. Instead, fluid tissues increase their cell shape index, and then we expect the condition $\beta k_E>k_C$, with $\beta>1$.

If we now consider a two-dimensional system as in Fig.~\ref{fig.FigS4}B-(left) and activate the red junction, the system will decrease its total junction length ($\beta <1$). Instead, if we consider a two-dimensional system as in Fig.~\ref{fig.FigS4}B-(right), with the condition that outer ends of the shoulder junctions can move vertically while keeping $2l_2+l_1=3L$ fixed, we get 
\begin{align}
\delta L^- &=L-l_1,\\
\delta L^+ &= 4(l_2-L) = 4(3L/2 -l_1/2 -L)= 2(L-l_1).
\label{eq.2Dcase}
\end{align}
Thus, $\beta=\delta L^+/\delta L^- = 2$.

\section{Additional characterizations of tissue mechanics from the vertex model simulations}

\begin{figure}[ht!] 
\centering 
 \includegraphics[width=1\linewidth]{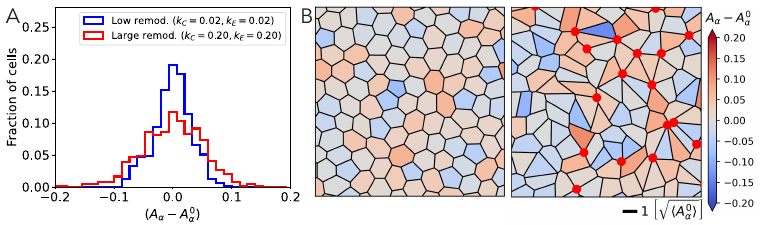}
\caption{\textbf{Role of tension remodeling on the distribution of cellular pressure}. (A) Histogram of cellular pressure for low $(k_C/k_L=k_E/k_L=0.02)$ and higher $(k_C/k_L=k_E/k_L=0.20)$ rates of tension remodeling. (B) Snapshots of the tissues considered in (A) show the spatial distribution of cellular pressure, for low (left) and high (right) rates of tension remodeling. Red circles represent four-fold vertices.}
\label{fig.FigS5}
\end{figure}

\noindent {\bf Effect of tension remodeling on cellular pressure}. Figure \ref{fig.FigS5} shows the distribution of cellular pressure in the tissue, which is given for each cell as $K (A_\alpha-A_\alpha^0)$. The pressure distribution becomes wider and larger in magnitude for higher rates of tension remodeling. This suggests that pressure-like forces play an important role in regulating tissue topology and the stability of four-fold vertices.\\

\begin{figure}[ht!] 
\centering 
 \includegraphics[width=.7\linewidth]{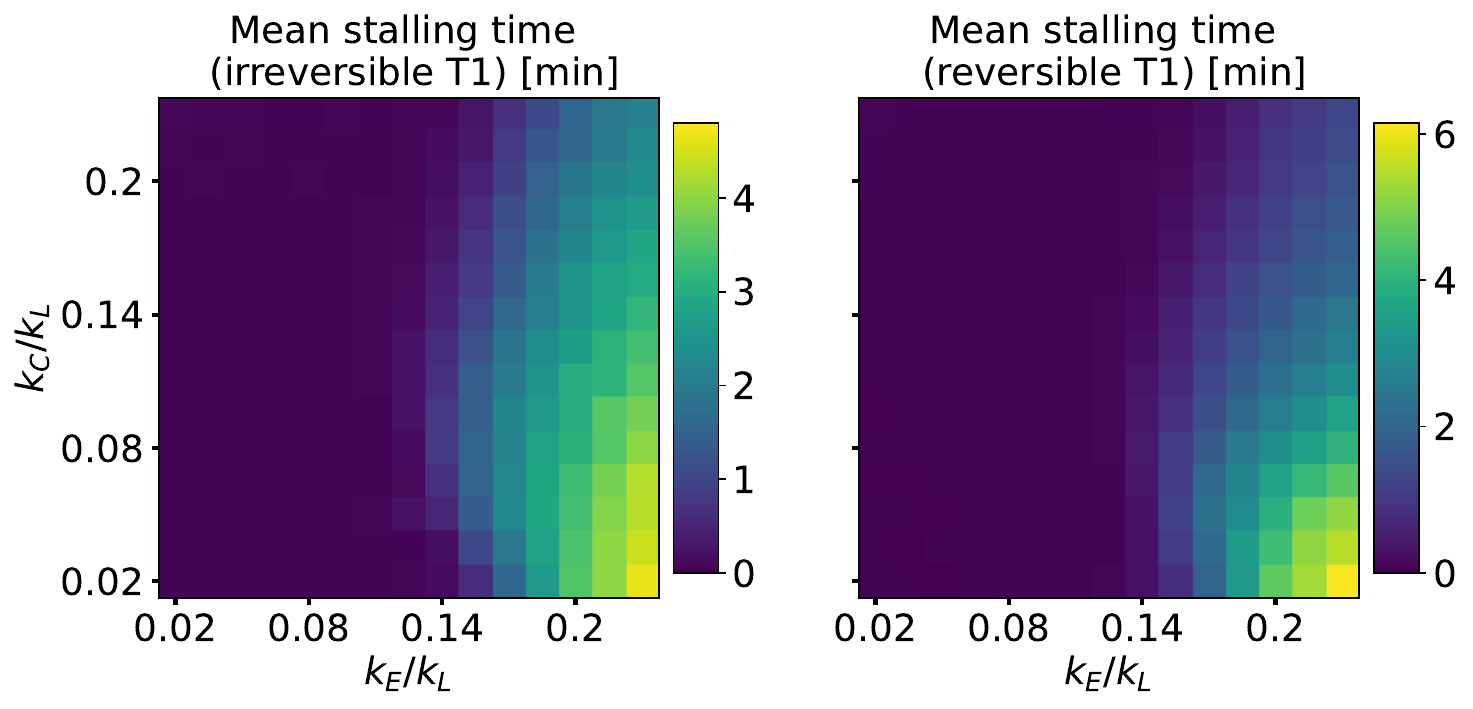} 
\caption{Mean T1 stalling time (colorscale) as functions of normalized tension remodeling rates, $k_E/k_L$ (x-axes) and $k_C/k_L$ (y-axes). Left: Mean stalling time for irreversible T1 transitions. Right: Mean stalling time for reversible T1 transitions.}
\label{fig.FigS6}
\end{figure}

\noindent {\bf Effect of tension remodeling on T1 stalling times}. Figure \ref{fig.FigS6} shows the role of tension remodeling on the mean stalling time (in minutes) for T1 transitions. Four-fold vertices are present for longer times for large $k_E$ and small $k_C$, reaching mean stalling times of 4.5 min and 6 min, for T1 and reversible T1 events, respectively.\\

\noindent {\bf Correlation between T1 stalling time and T1 rates}. Figure \ref{fig.FigS17} shows the correlation between the rate of T1 transitions per junction and the mean T1 stalling time. The initial increase in the rate of T1 with T1 stalling times indicates that stable four-fold vertices are present in fluid-like tissues with a high rate of T1 events. A negative correlation between the mean stalling time and the rate of T1 events emerges for higher stalling times. The presence of four-fold vertices for longer times implies a decrease in the events of cellular rearrangements.\\

\noindent {\bf Role of tension noise amplitude $\sigma$}. In the main text, all the simulations considered a fixed amplitude of tension fluctuations, $\sigma=0.02$.
Fig.~\ref{fig.FigS13}A shows how the rate of instantaneous and delayed events change when using different values of the tension noise amplitude $\sigma$. We find that tension fluctuations increase the number of T1 events. Additionally, we obtain that T1 stalling time decreases with increasing tension fluctuations (Fig.~\ref{fig.FigS13}B). Stable four-fold vertices can be present for more than 40 minutes if fluctuations are minimal. \\

\noindent {\bf Role of tension relaxation $\tau_\Lambda$}. In the main text, all simulations were run considering a fixed value for the tension relaxation timescale $\tau_\Lambda=10$. 
Fig.~\ref{fig.FigS14}A shows how the rate of instantaneous and delayed events change when using different values of $\tau_\Lambda$. Larger tension relaxation timescales decrease the number of T1 events, while increasing T1 stalling times (Fig.~\ref{fig.FigS14}B). Fig.~\ref{fig.FigS18} shows the temporal evolution of tissue mechanical energy for $(k_C,k_E)=(0.02,0.23)$ (left), and $(k_C,k_E)=(0.23,0.02)$ (right), while varying $\tau_\Lambda$. We find that $\tau_\Lambda$ sets the steady-state energy of the tissue. Furthermore, increasing $\tau_\Lambda$ decreases the threshold for transition from solid-to-fluid phase. Fig.~\ref{fig.FigS19} shows the comparison between the phase diagrams obtained for $\tau_\Lambda=10$ and $\tau_\Lambda=20$. \\

\noindent {\bf Finite shear simulations over solid and fluid tissues}. We perform external finite shear over tissues defined as solid and fluid by cell's center diffusion, with (Fluid II) and without (Fluid I) transiently stable 4-fold vertices, Fig.~\ref{fig.FigS16}. The relaxation scale for the solid tissue is of hours, while for the fluid tissues is of minutes. Tissue Fluid-II releases energy and reduces stress faster than Tissue Fluid-I, even though it has fewer instantaneous T1 events, or instantaneous cell neighbor exchanges. 
We find that active tension remodeling in addition to mechanical memory loss mechanisms develop transiently stable rosettes in  Fluid-II, generating a mechanical material that responds more liquid-like than Fluid-I.\\

\section{Analysis of different rules during the creation and resolution of 4-fold vertices}

\begin{figure}[ht!] 
\centering 
 \includegraphics[width=.5\linewidth]{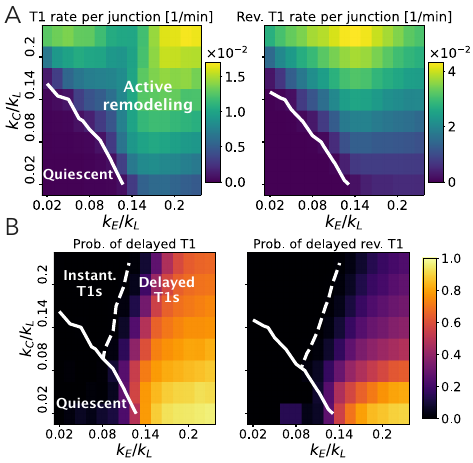} 
\caption{\textbf{Phase diagrams with no critical strain ($\epsilon_c=0$) for tension remodeling}. (A) Rates of T1 (left) and reversible (right) transitions for different values of $k_E/k_L$ and $k_C/k_L$. Solid lines represent $10^{-3}$ T1 events per junction per minute. (B) Probability of stalled/delayed irreversible T1 transitions (left) and reversible T1 events (right), for different values of $k_E/k_L$ and $k_C/k_L$. Dashed lines represent $1\%$ probability.
}
\label{fig.FigS7}
\end{figure}

\noindent {\bf Effect of varying the critical strain parameter $\epsilon_c$.} The threshold strain for tension remodeling is motivated by experimental data on single junction activation. However, we can remove this parameter from our model Fig.~\ref{fig.FigS7} by setting $\epsilon_c=0$. We find that the absence of threshold strain does not affect the role of junction tension remodeling in controlling the rates of T1 transitions. Additionally, we find that varying the value of $\epsilon_c$ does not induce major changes in the probabilities of delayed events nor the distribution of T1 stalling times, Fig.~\ref{fig.FigS12}.\\

\begin{figure}[ht!] 
\centering 
 \includegraphics[width=.51\linewidth]{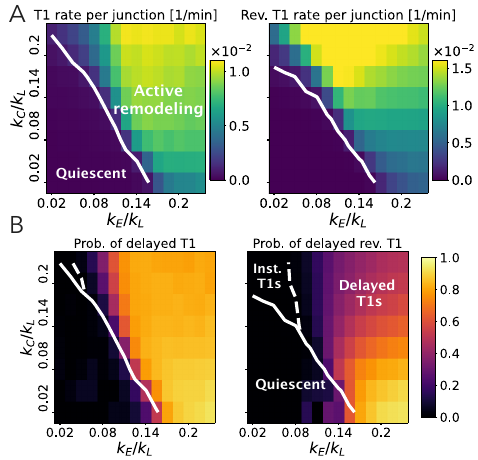} 
\caption{\textbf{Phase diagrams with a constant tension after T1 transition, $\Lambda_{\text{birth}}=\Lambda_0$.} (A) Rates of T1 (left) and reversible (right) transitions for different values of $k_E/k_L$ and $k_C/k_L$. Solid lines represent $10^{-3}$ T1 events per junction per minute. (B) Probability of stalled/delayed irreversible T1 transitions (left) and reversible T1 events (right), for different values of $k_E/k_L$ and $k_C/k_L$. Dashed lines represent $1\%$ probability. For these simulations we consider $\Lambda_{\text{birth}}=\Lambda_0$ (no noise on the tension of the newly born junction).
}
\label{fig.FigS8}
\end{figure}

\begin{figure}[ht!] 
\centering 
 \includegraphics[width=.5\linewidth]{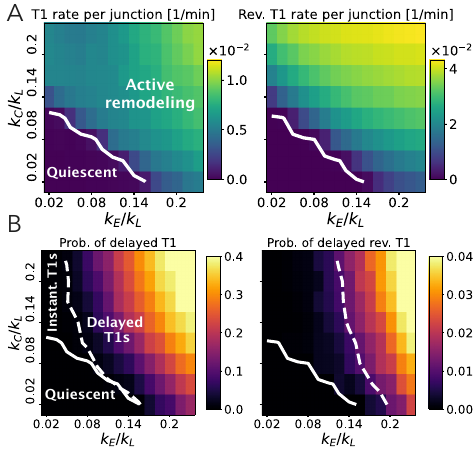} 
\caption{\textbf{Phase diagrams with a reduced tension after T1 transition, $\Lambda_{\text{birth}}=0.1\Lambda_0$.} (A) Rates of T1 (left) and reversible (right) transitions for different values of $k_E/k_L$ and $k_C/k_L$. Solid lines represent $10^{-3}$ T1 events per junction per minute. (B) Probability of stalled/delayed irreversible T1 transitions (left) and reversible T1 events (right), for different values of $k_E/k_L$ and $k_C/k_L$. Dashed lines represent $1\%$ probability. For these simulations we consider no change in shoulder junctions tension during the creation and resolution of 4-fold vertices, and $\Lambda_{\text{birth}}=0.1 \Lambda_0$ (no noise on the small tension of the newly born junction).
}
\label{fig.FigS9}
\end{figure}

\begin{figure}[ht!] 
\centering 
 \includegraphics[width=\linewidth]{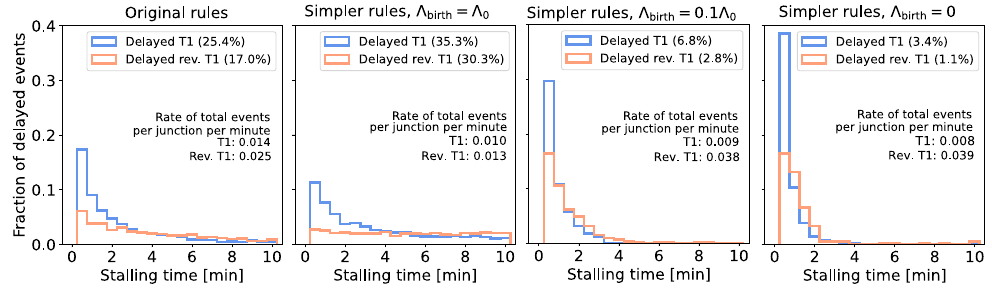} 
\caption{Comparison of histograms of the stalling time for delayed irreversible T1 (blue) and delayed reversible T1 (red) events, for $k_C/k_L=0.17,k_E/k_L=0.20$, when considering different tension resetting rules after a T1 transition. Simpler rules: no change in shoulder junctions tension during the creation and resolution of 4-fold vertices, and zero birth junction strain ($l_{\text{birth}}^0 = l_{\text{birth}}=1.5l_{T1}$).
}
\label{fig.FigS10}
\end{figure}

\noindent {\bf No change in shoulders tension and uniform $\Lambda_{\text{birth}}$.} In the main text, all simulations were run considering a gain and loss in the shoulders  tension during the creations and resolution of 4-fold vertices, respectively, motivated by experimental observations of Myosin-II accumulation around junctions proximal to 4-fold vertices. Here we present simulations results assuming no change in the shoulder tension, and $l_{\text{birth}}^0 = l_{\text{birth}}=1.5l_{T1}$ (simpler rules). First we use  $\Lambda_{\text{birth}}=\Lambda_0$ (mean initial tension), Fig.~\ref{fig.FigS8}. We obtain the same qualitative results (three regions: quiescent, instantaneous events, delayed events), with larger probabilities of transiently stable 4-fold vertices. Then, considering a small new tension $\Lambda_{\text{birth}}=0.1\Lambda_0$, Fig.~\ref{fig.FigS9} we obtained the same previous results. However, differently from the original simulations, under these rules the probability of stable 4-fold vertices increases with both $k_C/k_L$ and $k_E/k_L$. Particularly for $k_C/k_L=0.17, k_E/k_L=0.20$, we quantified the stalling times associated to delayed T1 and reverse T1 events for the different models, including the case of $\Lambda_{\text{birth}}=0$, Fig.~\ref{fig.FigS10}. Interestingly, even for $\Lambda_{\text{birth}}=0$ we obtain delayed events. However, we find that the probability and stalling time decrease with smaller $\Lambda_{\text{birth}}$.
 \\

\begin{figure}[ht!] 
\centering 
 \includegraphics[width=\linewidth]{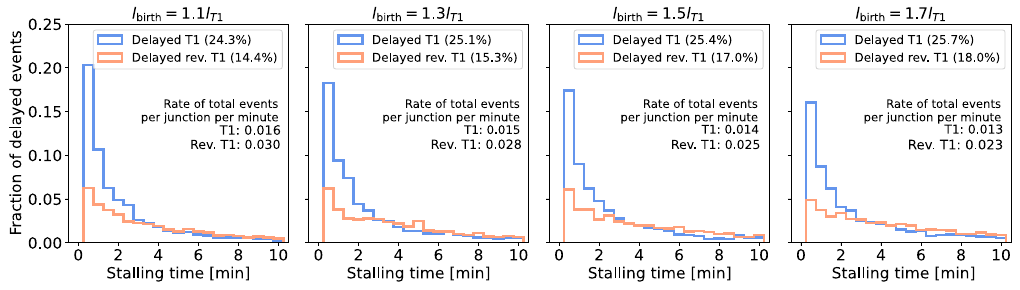} 
\caption{Comparison of histograms of the stalling time for delayed irreversible T1 (blue) and delayed reversible T1 (red) events, for $k_C/k_L=0.17,k_E/k_L=0.20$, when considering different values $l_{\text{birth}}$ (junction length after a T1 transition). 
}
\label{fig.FigS11}
\end{figure}

\begin{figure}[h!] 
\centering 
 \includegraphics[width=\linewidth]{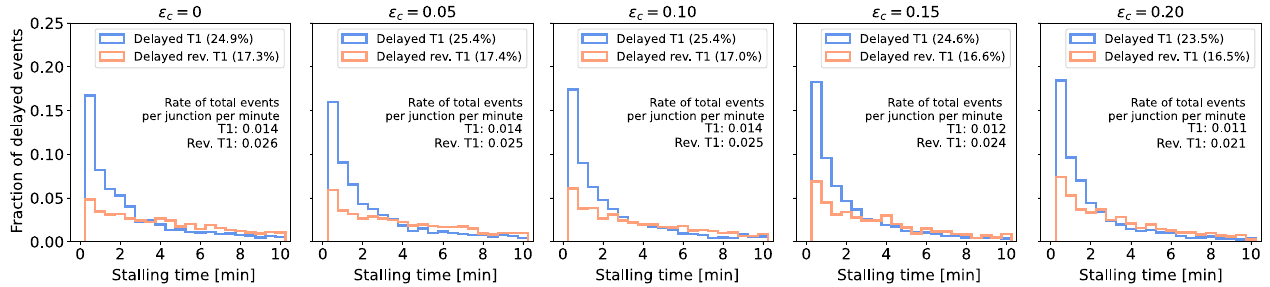} 
\caption{Comparison of histograms of the stalling time for delayed irreversible T1 (blue) and delayed reversible T1 (red) events, for $k_C/k_L=0.17,k_E/k_L=0.20$, when considering different values of the critical strain threshold, $\epsilon_c$. 
}
\label{fig.FigS12}
\end{figure}

\begin{figure}[h!] 
\centering 
 \includegraphics[width=.7\linewidth]{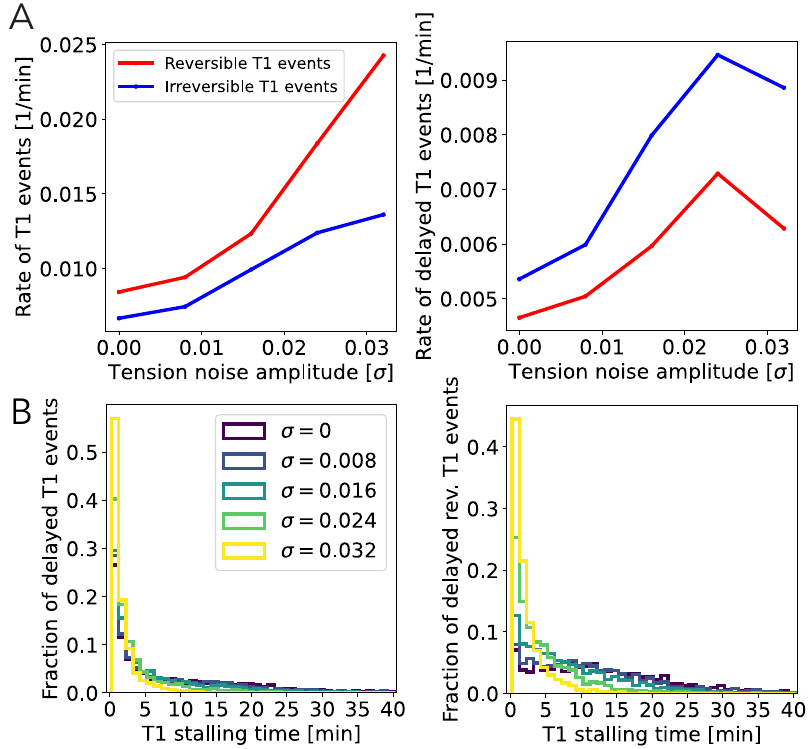} 
\caption{\textbf{Role of tension fluctuations on the rate of T1 events and T1 stalling times}. (A) Rate of irreversible and reversible T1 events (left: all T1 events, right: only delayed T1 events), for $(k_C/k_L=0.1,k_E/k_L=0.2)$ (fluid tissue) and different values of the tension noise amplitude $\sigma$. (B) Histogram of T1 stalling times for delayed irreversible T1 events (left) and delayed reversible T1 events (right), for different values of the tension noise amplitude $\sigma$.}
\label{fig.FigS13}
\end{figure}

\begin{figure}[ht!] 
\centering 
 \includegraphics[width=.75\linewidth]{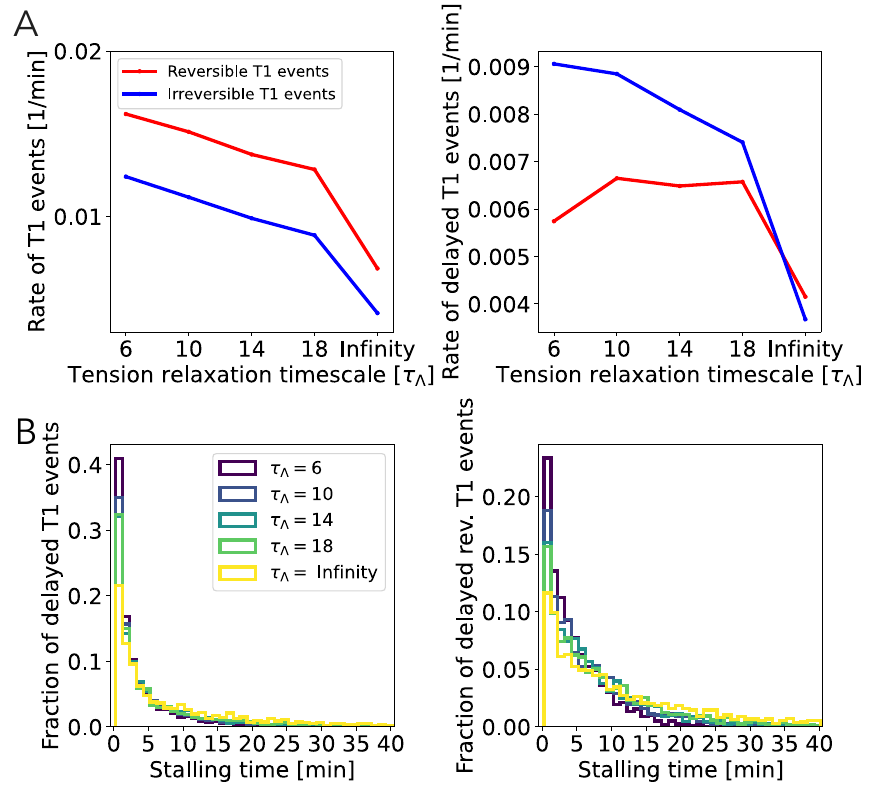} 
\caption{\textbf{Role of the tension relaxation timescale on the rate of T1 events and T1 stalling times}. (A) Rate of irreversible and reversible T1 events (left: all T1 events, right: only delayed T1 events), for $(k_C/k_L=0.1,k_E/k_L=0.2)$ (fluid tissue) and different values of the tension relaxation timescale $\tau_\Lambda$. (B) Histogram of T1 stalling times for delayed irreversible T1 events (left) and delayed reversible T1 events (right), for different values of $\tau_\Lambda$.}
\label{fig.FigS14}
\end{figure}

\noindent {\bf Effect of different $l_{\text{birth}}$ on T1 and reverse T1 stalling times.} In the main text, all simulations were run considering that right after resolution a new junction has a length $l_{\text{birth}}=1.5 l_{T_1}$. We run simulations for $k_C/k_L=0.17, k_E/k_L=0.20$, considering $l_{\text{birth}}/l_{T_1}=\lbrace{1.1,1.3,1.5, 1.7\rbrace}$. We do not observe any major changes on the probabilities of delayed events nor on the stalling times, Fig.~\ref{fig.FigS11}.   \\

\begin{figure}[ht!] 
\centering 
 \includegraphics[width=.5\linewidth]{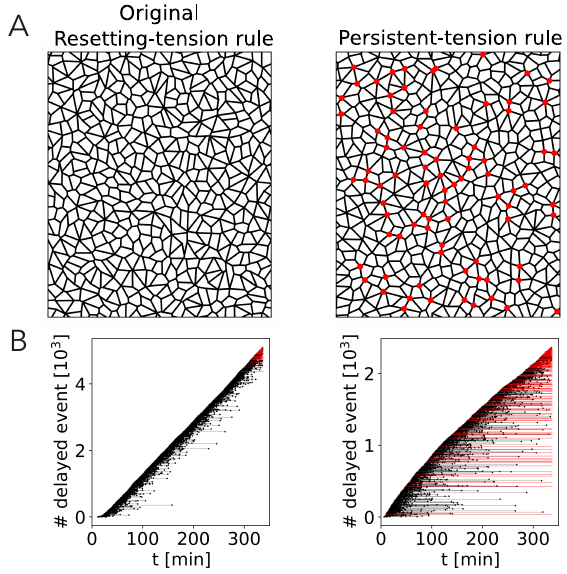} 
\caption{Comparison of T1 resolution times for delayed T1 events, for an active tissue with $k_C /k_L=0.1$, $k_E /k_L=0.20$, with (left column) tension-resetting model and (right column) a persistent tension model  during 4-fold vertex resolution. (A) Tissue configurations showing the steady-state morphology (at $\sim \SI{350}{\min}$), where red solid circles represent 4-fold vertices that have been stable for more than $\SI{100}{\min}$ by the end of each simulation. (B) Each line represents the creation of a 4-fold vertex. Black lines represent 4-fold vertices that are resolved through the simulation (at the time highlighted by a black dot), in a time larger than 6 seconds. Red lines, that finish in an empty red circle, represent 4-fold vertices that are not resolved during the simulation.}

\label{fig.FigS15}
\end{figure}

\begin{figure}[ht!] 
\centering 
 \includegraphics[width=.9\linewidth]{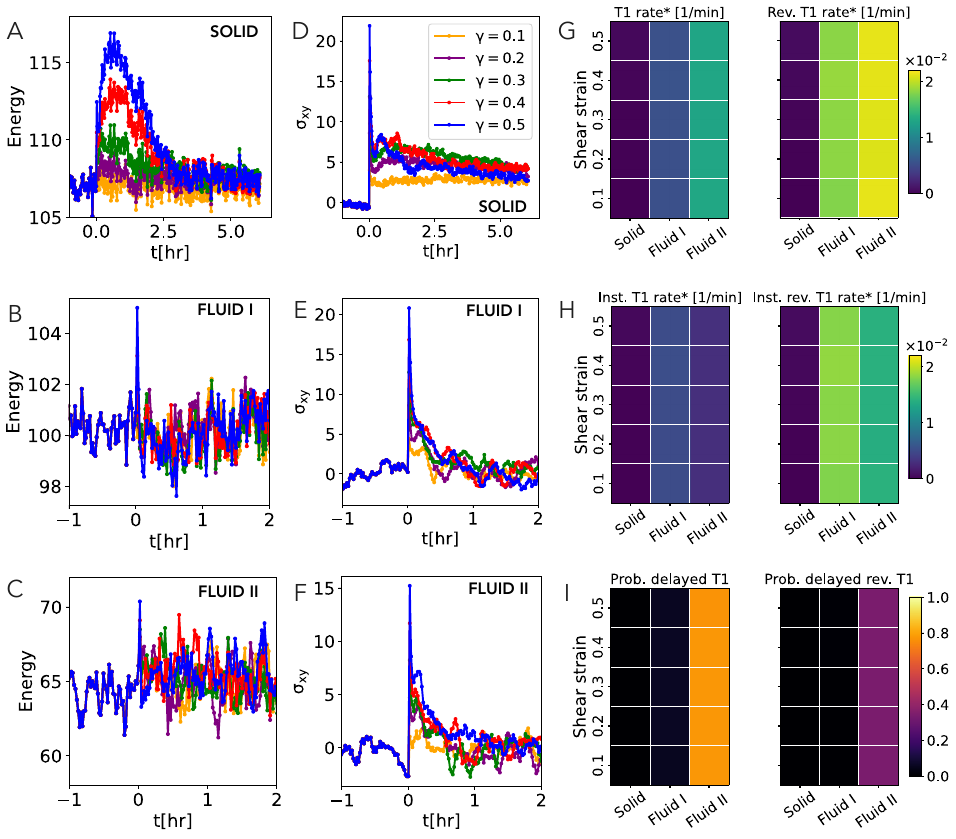} 
\caption{\textbf{Tissue response upon finite shear simulations}. Mechanical response (A-C: energy and D-F: stress release) under finite shear $\gamma = \lbrace{0.1,0.2,0.3,0.4,0.5\rbrace}$, applied at $t=0$, considering Edwards periodic boundary conditions, on three tissues with $k_C/k_L=0.14$: Solid ($k_E/k_L=0.05$), Fluid I ($k_E/k_L=0.11$, without transiently stable 4-fold vertices), and Fluid II ($k_E/k_L=0.20$, with transiently stable 4-fold vertices). Here, energy is defined as $E_{\text{el}} + \left(T_{ij} + \Gamma_a l_{ij}/2\right)l_{ij}$, and stress as $\sigma_{xy}=\sum_{i,j} l_{ij}^x l_{ij}^y \left(T_{ij} + \Gamma_a l_{ij}\right)/l_{ij}$. G: rate of T1 and reverse T1 events ($^*$ means normalized by 1482 junctions). H: rate of instantaneous events ($^*$ means normalized by 1482 junctions). I: Probability of delayed events. 
}
\label{fig.FigS16}
\end{figure}

\begin{figure}[ht!] 
\centering 
 \includegraphics[width=.5\linewidth]{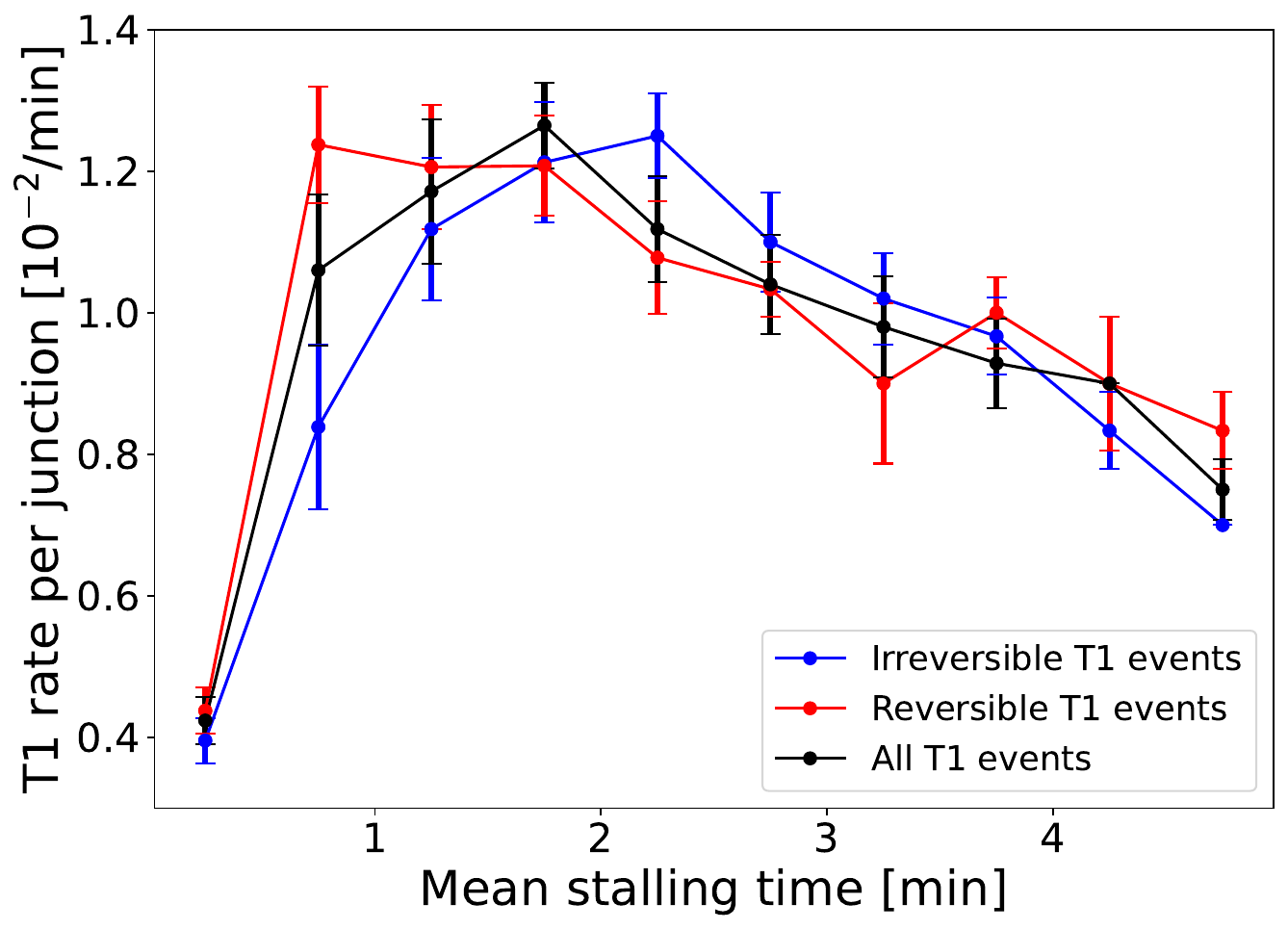} 
\caption{Correlation between the rate of T1 transitions and the mean T1 stalling times. A negative correlation emerges at larger stalling times. Error bars represent $\pm$1 standard error of mean.}
\label{fig.FigS17}
\end{figure}

\begin{figure}[ht!] 
\centering 
 \includegraphics[width=\linewidth]{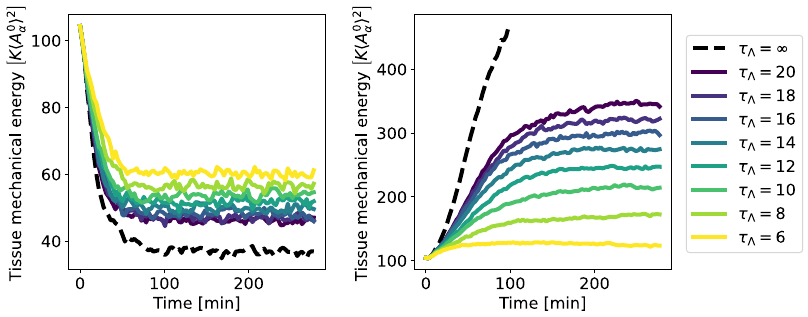} 
\caption{\textbf{Role of the tension relaxation timescale on the mechanical energy of the tissue}. Temporal evolution of the tissue mechanical energy, here defined as $E_{\text{el}} + \left(T_{ij} + \Gamma_a l_{ij}/2\right)l_{ij}$, for various values of the tension relaxation timescale $\tau_\Lambda$, corresponding to two different combinations of the tension remodeling rates, $(k_C/k_L=0.02,k_E/k_L=0.23)$ (left) and $(k_C/k_L=0.23,k_E/k_L=0.02)$ (right). Dashed black curves correspond to simulations with no tension relaxation, i.e. $\tau_\Lambda \rightarrow \infty$.}
\label{fig.FigS18}
\end{figure}

\begin{figure}[ht!] 
\centering 
 \includegraphics[width=.75\linewidth]{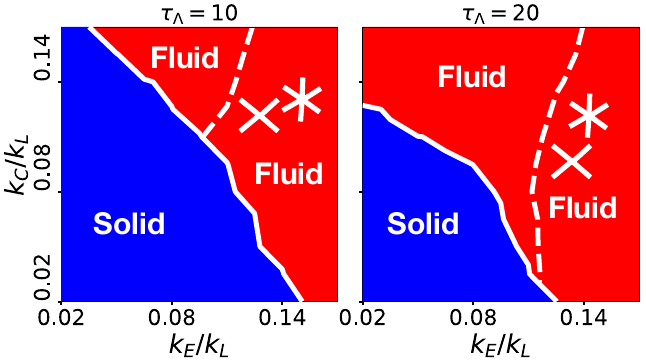} 
\caption{\textbf{Role of the tension relaxation timescale on solid-fluid phase transitions}. Phase diagram showing transitions between solid (blue) and fluid (red) states of the tissue, with solid white curve representing the phase boundary. Left:$\tau_\Lambda=10$. Right: $\tau_\Lambda=20$. Below the dashed white curve in the fluid phase, stable 4-fold vertices are prevalent.}
\label{fig.FigS19}
\end{figure}

\noindent{\bf Effect of persistent-tension rule.} We ran a simulation ($k_C/k_L=0.1, k_E/k_L=0.2$) considering a persistent-tension rule, i.e., a newly birth junction recovers the tension it used to have before forming the 4-fold vertex. We find the emergence of permanently stable 4-fold vertices in this case, as shown in Fig.~\ref{fig.FigS15}.   \\

\clearpage 
\bibliographystyle{ieeetr}
\bibliography{supplemental.bib}